\documentstyle[epsf]{l-aa}
\def\Tpc{T_{\rm pc}}
\def\Rpc{R_{\rm pc}}
\def\Tc{T_{\rm core}}
\def\Rc{R_{\rm core}}
\def\Tr{T_{\rm rim}}
\def\Rr{R_{\rm rim}}
\def\psr{PSR J0437--4715}
\def\fp{f_{\rm p}}

\def\rns{R_{\rm NS}}
\def\mns{M_{\rm NS}}
\def\be{\begin{equation}}
\def\ee{\end{equation}}
\begin{document}
\thesaurus{02.          
              (08.14.1;   
               08.16.7;   
               02.18.6;   
               13.25.5)   
            }
\title{ Soft X-rays from polar caps of
the millisecond pulsar J0437--4715}
\author{V.~E.~Zavlin\inst{1,}
\thanks{e-maill: zavlin@mpe-garching.mpg.de} and G.~G. Pavlov\inst{2,1}}
\institute{Max--Planck--Institut f\"ur Extraterrestrische Physik,
Giessenbachstrasse, D-85740 Garching, Germany \and
Pennsylvania State University, 525 Davey Lab, University Park,
PA 16802, USA}
\date\today  
\maketitle
\markboth{Zavlin and Pavlov: X-rays from polar caps of \psr}{}
\begin{abstract}
We show that the soft X-ray spectra and light curves 
observed with the $ROSAT$ and $EUVE$ from  
the closest known millisecond pulsar J0437--4715 
can be interpreted as thermal radiation from two hot polar caps 
whose emitting layers (atmospheres) are comprised of hydrogen.
The simplest model yields a uniform temperature of $(0.8-0.9)\times 10^6$~K 
within a cap radius of $0.7-0.9$~km. The spectral fits indicate that the 
temperature may be nonuniformly distributed along the cap surface.
The distribution can be approximated by a central core heated up to
$(1-2)\times 10^6$~K within a radius of $0.2-0.4$~km,
surrounded by a colder rim with temperatures
$(3-5)\times 10^5$~K extending out to $2-6$~km. 
The polar cap interpretation implies low column densities, $(1-3)\times
10^{19}$~cm$^{-2}$, and a high degree of ionization,
$> 20$\%, of the interstellar hydrogen towards the pulsar.
The inferred bolometric luminosity of the polar caps,
$(1.0-1.6)\times 10^{30}$~erg~s$^{-1}$, is in excellent 
agreement with the predictions of the slot-gap model
of radio pulsars developed by Arons and his coworkers.
Similar polar cap radiation should be emitted by other
millisecond pulsars, although in some of them (e.~g., PSR B1821--24)
the soft X-ray flux is dominated by the nonthermal radiation
from pulsar magnetospheres.

\end{abstract}

\newcommand{\gapr}{\raisebox{-.6ex}{\mbox{
$\stackrel{>}{\mbox{\scriptsize$\sim$}}\:$}}}
\newcommand{\lapr}{\raisebox{-.6ex}{\mbox{
$\stackrel{<}{\mbox{\scriptsize$\sim$}}\:$}}}

\section{Introduction}
The bright 5.75~ms pulsar J0437--4715
was discovered by Johnston et al.~(1993) during 
the Parkes survey for millisecond pulsars.
The pulsar is in a 5.74~d binary orbit with a cool 
($T_{\rm color}\simeq 4000$~K), low-mass ($\sim 0.2 M_\odot$) white
dwarf companion and is surrounded by a bow-shock
nebula (Becker et al.~1993; Bailyn 1993; Bell, Bailes \& Bessel 1993; Danziger,
Baade \& Della Valle 1993; Bell et al.~1995). 
It is an old object, with a characteristic age $\tau=
P/2\dot{P}\simeq 5\times 10^9$~yr, low magnetic field
$B\sim 3\times 10^8$~G, and rotational energy loss
$\dot{E}=4\times 10^{33}$~erg~s$^{-1}$.
The observed dispersion measure of 2.65~pc~cm$^{-3}$ implies a distance
$d\simeq 100-180$~pc, making this the closest known millisecond pulsar. 
Sandhu et al.~(1997) reported the distance $d=178\pm 26$~pc,
from parallax measurements.
The pulsar shows significant radio emission over at least 80\%
of the pulse period, with a complicated mean pulse shape varying with
radio frequency. Variation  of
the linear polarization position angle within the mean pulse
interpreted in terms of the rotating vector model
(Radhakrishnan \& Cooke 1969) yields the angle between
the observer's line of sight and the rotation axis, $\zeta\simeq 40^\circ$,
and the angle between the magnetic and rotation axes,
$\alpha\simeq 35^\circ$ (Manchester \& Johnston 1995).

$ROSAT$ observations of \psr~ with the Position
Sensitive Proportional Counter (PSPC) have revealed
(Becker \& Tr\"umper 1993; hereafter BT93) that this
is also a bright (count rate $=0.204\pm 0.006$~s$^{-1}$)
soft X-ray pulsar with a single broad pulse and a pulsed
fraction $\fp =33\pm 3\%$ in the PSPC energy range $0.1-2.4$~keV.
BT93 found that the pulsed fraction varies with photon energy $E$ and peaks
in the range $0.6-1.1$~keV, reaching $53\pm 6\%$.
Although the pulsar spectrum can be fitted with a single
power law, 
indicative of a non-thermal origin of the soft X-ray radiation, 
the energy dependence of the pulsed fraction in the narrow
energy range makes this interpretation hardly plausible.
BT93 showed also that a single blackbody model does not
fit the PSPC spectrum leaving a residual hard excess above
0.4~keV. They suggested that the spectrum can consist of
two components: a power law, representing magnetospheric
or nebular emission, 
and a blackbody component of the 
temperature $T\sim 1.7\times 10^6$~K 
emitted from an area of $\sim 0.08\, d_{180}^2~{\rm km}^2$ 
($d_{180}=d/180~{\rm pc}$); this thermal component
was suggested by BT93 to be radiated
from a hot spot on the neutron star (NS) surface.

\psr~ was also detected with the $EUVE$ Deep Survey Instrument
(DSI) in the Lexan band filter ($E\simeq 0.05-0.2$~keV) by
Edelstein, Foster \& Bowyer (1995) and Halpern, Martin \& Marshall (1996;
hereafter HMM96). According to Edelstein et al.~(1995),
the DSI count rate of the source is $0.0143\pm 0.0008$~s$^{-1}$,
whereas HMM96 report the count rate of $0.00973\pm 0.00017$~s$^{-1}$
obtained with a much longer exposure (496~ks vs.~72~ks).
Edelstein et al.~(1995) ruled out 
the single power-law spectral model based on the
improbably high hydrogen column density, $n_H=2.5\times 10^{20}$~cm$^{-2}$,
required by this model. They claimed that both the 
$ROSAT$ flux below 0.4~keV and the
$EUVE$ flux could arise from an isothermal blackbody with
a temperature $\sim 5.7\times 10^5$~K, an emitting area of $\sim 3$~km$^2$,
and an absorbing column of $n_H=5\times 10^{19}$~cm$^{-2}$.
On the other hand, HMM96 concluded that the combined analysis
of the $ROSAT$ spectrum and $EUVE$ flux is consistent with
a single power-law spectrum of photon index $\gamma=2.2-2.5$ and
intervening column density $n_H=(5-8)\times 10^{19}$~cm$^{-2}$.
Alternatively, the combined data can be interpreted as comprised
of two components, e.~g., a power law and a blackbody component emitted
from a hot polar cap of radius 50--600~m and temperature
$(1.0-3.3)\times 10^6$~K. HMM96 observed the pulsar in the
high time resolution mode, which enabled them to obtain the
light curve and to measure the pulsed fraction $\fp = 27\pm 5\%$ in the
$EUVE$ DSI spectral range.

Thus, although very important data have been collected
and analyzed, the true nature of the soft X-ray radiation from \psr~ remains
elusive. The only firmly established facts are that the radiation
is pulsed (pulsed fraction apparently depends on energy), and the spectrum
cannot be fitted with a single blackbody model. The main question 
is whether the radiation is of a nonthermal (magnetospheric? nebular?) 
origin or at least a fraction of it can be interpreted
as thermal (or thermal-like) radiation from some heated regions
(polar caps?) on the NS surface.

Virtually all the (different) models of radio pulsars 
(e.~g., Cheng \& Ruderman 1980;
Arons 1981; Michel 1991; Beskin, Gurevich \& Istomin 1993)
predict these objects to have {\em polar caps} (PCs) around
the NS magnetic poles heated up to X-ray temperatures by
relativistic particles and gamma-quanta impinging onto the pole regions from
the acceleration zones. A conventional assumption about
the PC radius is that it is close to the radius within which
open magnetic field lines originate from the NS surface. For \psr, it gives
$\Rpc =1.9\, (\rns /10\,{\rm km})^{3/2}\,(P/5.75\,{\rm ms})^{-1/2}$ km. 
Expected PC temperatures, $T_{\rm pc}\sim 5\times 10^5 - 5\times 10^6$~K, 
and luminosities, $L_{\rm pc} \sim 10^{28}-10^{32}$~erg~s$^{-1}$,
are much less certain, being strongly
dependent on the specific pulsar model. Thus, 
one cannot firmly predict
PC properties because of the lack of a well-established
pulsar model --- rather a theoretical model should be chosen based
on X-ray observations of radio pulsars.

X-ray observations of
other old pulsars (e.~g., PSR B1929+10 -- Yancopoulos, Hamilton \&
Helfand 1994; PSR B0950+08 -- Manning \& Willmore 1994) do show pulsed
X-ray radiation which, in principle, could be the thermal PC
radiation. However, the number of photons collected has been too
small to make firm conclusions, and the opposite hypothesis,
that this radiation is of a magnetospheric origin (\"Ogelman 1995;
Becker \& Tr\"umper 1997 -- hereafter BT97), cannot be excluded.
In principle, one could also observe the PC radiation from
younger pulsars and use these data to discriminate between
different PC models. Indeed, there are some indications that
hard components of the soft X-ray spectra of, e.~g., PSR B0656+14
and PSR B1055--52 (Greiveldinger et al.~1996) may consist
of two subcomponents, a power law and a thermal subcomponent corresponding
to emission from PCs of temperatures $\Tpc \sim 1.5\times 10^6$~K (for
PSR B0656+14) and $\sim 3.7\times 10^6$~K (for PSR B1055--52).
This interpretation, however, is not
unique because it is difficult to separate the hard component
from the soft one which is believed to originate from the
(cooler) entire NS surface, and even more difficult
to separate the two subcomponents of the hard spectral tail.

One cannot also exclude {\em a priori} that the nonthermal
radiation from the pulsar magnetosphere or a pulsar-powered
compact nebula contributes to, or
even dominates, the observed soft X-ray flux of \psr.
It is natural to assume that the spectrum of this radiation 
can be approximated by a power law in the relatively
short $EUVE$-$ROSAT$ range. One could also expect the magnetospheric
(but not nebular) radiation to be pulsed with the
radio pulsar period; the pulses should be, 
as a rule, narrower
than those of the thermal PC radiation, and the shape of the light
curve should not vary considerably with photon energy.
Thus, if more data confirm the conclusion of BT93 that
the pulsed fraction depends on $E$, the radiation of \psr~
should either be thermal or consist of the thermal and nonthermal
components. In the latter case, since the sources
of the thermal and nonthermal radiation are expected to be
spatially separated, it is natural to expect that pulsations 
of these two components should be phase-shifted due to the
difference of travel times and aberration. For instance,
if the nonthermal component is generated at a distance
comparable to the light cylinder radius 
($R_{\rm lc}\sim Pc/2\pi \sim 3\times 10^7$ cm), the time delay
leads to a phase shift of about 0.2 of the pulsar period.
No energy-dependent phase shifts have been reported for this source,
although it can be explained by the poor photon statistics
in hard PSPC channels.

As in the case of thermal PC radiation, the current theoretical models
of nonthermal high-energy pulsar emission (e.~g.,
Sturner, Dermer \& Michel 1995; Romani 1996) are not enough
elaborated to predict the intensity, spectrum and light curve
for a given pulsar. There exist empirical formulae
(Seward \& Wang 1988; \"Ogelman 1995; BT97) which 
relate the (presumably nonthermal)
X-ray luminosity to the pulsar parameters,
e.~g., to the period and magnetic field. For instance,
\"Ogelman (1995) pointed out that the observed
soft X-ray luminosities satisfy, for 
7 pulsars, the following
equation: $L_x\simeq 6.6\times 10^{26}\, (B_{12}/P^2)^{2.7}$~erg~s$^{-1}
\propto \dot{E}^{1.35}$. 
BT97 fitted the luminosities of 26 pulsars observed
in the $ROSAT$ range with the dependence $L_x \simeq 0.001 \dot{E}$.
These pulsars represent a very wide range of
spin-down luminosities ($10^{33}-10^{39}$~erg~s$^{-1}$), ages 
($10^3-7\times 10^9$~yr), magnetic field strength ($10^8-10^{13}$~G) and
spin periods ($1.6-530$~ms). The inferred dependence
indicates that for most pulsars
the bulk of observed radiation is directly connected with
pulsar mechanisms, i.~e., with production and acceleration of
relativistic particles which carry away the NS rotational energy. 
The fact that X-rays from some of these
pulsars (e.~g., Crab) are certainly nonthermal may  allow
one to assume that the radiation detected with
the $ROSAT$ from most pulsars is 
of a nonthermal origin.
On the other hand, a correlation between $L_x$ and
$\dot{E}$ should take place also for radiation emitted
by the pulsar PCs because their (thermal) luminosity
is also provided by relativistic particles generated
in the magnetosphere and accelerated towards the NS surface,
so that the PC luminosity should be a fraction of $\dot{E}$
which goes to heating of the PCs. Moreover, both theoretical
(see Section 5.1) and observational estimates of the PC
luminosities may happen to be very close to the values predicted
by the $L_x(\dot{E})$ dependence obtained by BT97. For instance,
the luminosity of the thermal component in
the power-law plus blackbody fit for \psr,
$L_{\rm bol}=2.2\times 10^{30}\, d_{180}^2$~erg~s$^{-1}$ (BT93),
is close to the predicted value, $4\times 10^{30}$~erg~s$^{-1}$.
Another example is PSR B1929+10 whose luminosity
$L_{\rm bol}\simeq 1.2\times 10^{30}\, d_{250}^2$~erg~s$^{-1}$
inferred from the blackbody fit of both $ROSAT$ and $ASCA$ data
(Yancopoulos et al.~1994; Wang \& Halpern 1997) perfectly
matches the dependence derived by BT97.
In general case, we may expect that
the pulsar X-ray radiation
contains both thermal (PC) and nonthermal components, both
growing with $\dot{E}$, and the relation between the
thermal and nonthermal 
fluxes may be different for different objects, depending,
in addition to $\dot{E}$, on other
intrinsic pulsar parameters (e.~g., pulsar period $P$,
magnetic inclination $\alpha$), as well as
on the rotational inclination $\zeta$ (because
the radiation beam widths are different for the thermal
and nonthermal components). 
Thus, one cannot rule out that the PC
component dominates in some cases, and we explore this
possibility for \psr.

An important evidence on the nature of radiation
of \psr~ could come from deep observations of its
high-energy tail, at energies
above $1-2$~keV. Although this object has been
detected by $ASCA$ (two 20~ks observations) below 3~keV, the
results are still not very conclusive because of poor choice
of observing modes which did not take into account the
presence of a neighboring Seyfert galaxy. Nevertheless,
preliminary results show that a spectrum of $\sim 400$ pulsar
photons extracted from the CCD away from the Seyfert galaxy
is softer than a power law and resembles more a thermal-like
spectrum (Kawai, Tamura \& Saito 1996). This conclusion, however, needs
verification based on longer pointed observations in a mode
minimizing contamination from the Seyfert.

The conclusion of BT93 and HMM96 that the PSPC spectrum
of \psr~ cannot be fitted with a single thermal component
is based on the assumption that its spectrum coincides
with the spectrum of the blackbody radiation. However, 
spectra emitted by real bodies, including stars, are always different
from the Planck spectrum. In particular, if the temperature of
a stellar atmosphere grows inward, and the absorption coefficient
decreases with frequency (e.~g., $k_\nu \propto \nu^{-3}$ for
the Kramers law), then the spectrum is harder than the blackbody spectrum
at high frequencies because we see deeper and hotter layers.
This means that if a NS is covered with a fully ionized plasma,
whose opacity can be described by the Kramers law, its spectrum
is substantially harder at high energies, $h\nu \gapr kT_{\rm eff}$
(e.~g., Pavlov \& Shibanov 1978). This general property has been 
demonstrated in models of hydrogen and helium atmospheres of cooling NSs
by Romani (1987), Pavlov et al.~(1995), Rajagopal \& Romani~(1996),
Zavlin, Pavlov \& Shibanov (1996; hereafter ZPS96). One can expect
that a NS has a purely hydrogen atmosphere if it experienced
accretion of the interstellar matter --- heavy elements of the
accreted matter sink down rapidly due to the strong NS gravitation
(Alcock \& Illarionov 1980).
Since \psr~ is a very old object, it is quite plausible that
it accreted some matter during its long life, and in this case
we may expect that the excess of the observed flux at $E\gapr 0.4-0.6$~keV
is due to the fact that \psr, including its PCs, is covered
by a fully ionized hydrogen atmosphere.

The PC radiation should be inevitably pulsed unless the rotation axis
coincides with either the line of sight or the magnetic axis.
If it were the blackbody radiation, the shape of the light curves and the
pulsed fraction $\fp$ would remain the same at all photon energies $E$.
However, the atmosphere radiation has a very important feature
--- it is anisotropic, with anisotropy
(and consequently light curves) depending on energy (Pavlov et al.~1994; 
Shibanov et al.~1995; ZPS96).
This dependence is different for different chemical compositions and surface
temperatures. E.~g., for atmospheres with relatively
high temperatures $\sim 10^6$~K consisting of light
elements (hydrogen, helium), the anisotropy increases with increasing $E$
in the soft X-ray range.
This also makes the spectra dependent on the rotation phase.
The inter-dependence of the spectral and angular distributions 
means that the proper interpretation of the observed
PC radiation implies fitting of
{\em both the spectra and the light curves with the same PC model}.

The above-described properties of radiation emitted by a PC
covered with an atmosphere warrant a new investigation of
the soft X-ray radiation observed from \psr, based on the
NS atmosphere models. To perform this investigation, we used
the $ROSAT$ and $EUVE$ data analyzed previously
by BT93 and HMM96 and combined with the results of new observations
carried out with the $ROSAT$ High-Resolution Imager (HRI)
and $ROSAT$ PSPC (Becker et al.~1997). 
These observations are briefly described in Section~2.

We show that with allowance for the properties of  
NS atmosphere radiation these joint $ROSAT$ and $EUVE$ data can be
interpreted as thermal radiation of two PCs, {\em without invoking
an additional nonthermal component}. In Section~3
we explore a simplest single-temperature PC model which assumes
that both PCs have equal radii $\Rpc$
and uniform temperatures $\Tpc$, whereas the rest part
of the NS surface has a much lower temperature and is invisible
in X-rays. Since the magnetic field of \psr~ is very low,
it cannot affect the radiative properties of the PC atmospheres.
Therefore, we use the low-field atmosphere models developed
by ZPS96. To calculate the flux as measured by a distant observer, we
integrate the specific flux
over the visible PC surface with allowance for the gravitational
redshift and bending of the photon trajectories. 
For the case of small PCs ($\Rpc\ll\rns$)
with uniform temperatures, a convenient
expression for the observable flux
is given by Zavlin, Pavlov \& Shibanov (1995; see their Eq.~[A15])).
We show that this simple model is generally consistent with
the observational data if the PCs are covered with hydrogen
or helium, whereas the iron atmosphere model does not
fit the spectra (cf. Pavlov et al.~1996b).
For the hydrogen and helium models,
we estimate $\Rpc$, $\Tpc$ and the 
interstellar hydrogen column density, $n_H$. 

The fit with the simplest PC model is still not perfect.
This is not surprising because it is hard to expect the
real PCs to be uniformly heated. On the contrary,
due to higher heat conduction of subphotospheric layers,
where the energy of accreting relativistic particles is released,
the heat can propagate along the surface, eventually  heating
surface layers out of the ``primary'' hot spot. This should
result in a larger hot region 
with the temperature decreasing outward. 
In fact, this mechanism should be more efficient just for low-field 
pulsars because the strong magnetic fields
of ordinary pulsars greatly
reduce the transverse conductivity (e.~g., Hernquist 1985).
To the best of our knowledge, there have been no reliable
calculations of the temperature distribution around the pulsar
magnetic poles. Thus, to include this effect into consideration,
we explore in Section~4 a simple model:
the distribution is assumed to be a two-step
function (``core+rim'') with two temperatures, $\Tc$ and $\Tr$,
and two radii, $\Rc$ and $\Rr$. We show that this model
is fairly consistent, for hydrogen-covered PCs, with all the data available.  
We discuss the results and implications of our interpretation
in Section~5 and draw conclusions in Section~6.

\section{Observational data and the power-law fit}

\subsection{$ROSAT$ PSPC and HRI}
We used two data sets obtained during pointed PSPC
observations (total $3172\pm 58$ counts)
and one data set obtained with HRI
($1517\pm 39$ counts); the corresponding count rates are
$0.202\pm 0.004$~s$^{-1}$ and
$0.043\pm 0.004$~s$^{-1}$, respectively. Details of these
observations are described by BT93 and Becker et al.~(1997).

Since no source counts were detected in the PSPC pulse height channels above 
channel 200, and the PSPC response matrix is not
sufficiently known for softest channels, we used channels 10--200
($E\approx 0.1-2.0$~keV) for our spectral and temporal analysis. 
We manually binned the count spectrum 
into 49 energy bins with the signal-to-noise 
ratio between 4 and 9 and used this binning for the
spectral fitting with the aid of the 
MIDAS/EXSAS software (Zimmerman et al.~1994).

We used the PSPC and HRI temporal dependences 
of the source count rate folded with the radio pulsar period
by Becker et al.~(1997).
The PSPC light curves were extracted 
in 0.1--0.6, 0.6--1.1, 1.1--2.0 and 0.1--2.0~keV energy ranges
(13, 9, 9 and 17 phase bins, respectively).
The shapes of the 
light curves from the second data set were compared with those
of the first set using a $\chi^2$ test. 
The analysis showed that the light curves from both sets of the PSPC
data are consistent with each other with a 
probability $> 70\%$,
that justifies summation of the light curves from the two PSPC observations.

The combined PSPC light curves show an evidence that
the pulsed fraction $\fp$ depends on energy,
in accordance with the dependence reported by BT93 for
the first data set. 
Defining $\fp$ in the standard way,
as the fraction of the total counts lying above the light curve minimum,
we obtained an increase 
from $\fp =32\pm 4$\% at 0.1--0.6~keV to $\fp =44\pm 8$\% at 0.6--1.1~keV.
Formally, in the 1.1--2.0~keV range the pulsed fraction reaches 
$54\pm 11$\%, but this value should be considered with caution
because of poor statistics ($\simeq 252$ counts
in this range).
The pulsed fraction in the total energy range is $\fp =30\pm 4$\%.
No phase shifts between the light curves in different
energy ranges were found (cf.~BT93).

Although the estimated $\fp$ values indicate 
that the light curves have different shapes in different energy
ranges, they suffer from large uncertainties of the number
of counts in the phase bins corresponding to the minima of the observed
light curves, being thereby not very reliable estimators.
Besides, 
light curves with equal $\fp$ can have quite different shapes.
Therefore, an independent statistical test is needed
to ascertain and quantify statistically
the difference of the light curve shapes.
The same $\chi^2$ test shows that the probability
that the shapes of the light curves in different ranges
are {\em different} is generally high, between 77\% and 92\%. 
Although the difference might be partly due to systematic uncertainties,
we believe that the dependence of 
the light curve shapes and pulsed fractions on energy is 
statistically reliable (at least, it cannot be rejected
on statistical grounds) and can be considered as an additional argument 
in favor of thermal origin of the soft X-ray radiation from \psr.   

The HRI light curve folded with the radio pulsar 
period and binned into 11 phase bins 
results in the pulsed fraction $\fp=32\pm 5$\%.

\subsection{$EUVE$ DSI}
We used the observational data kindly provided by Jules Halpern
and described in detail by HMM96. The total of $4163
\pm 65$ source counts were collected 
at the source count rate of $0.00973\pm
0.00017$~s$^{-1}$. This uncertainty ($\pm 2\%$)
is due to statistics only; a systematic uncertainty may be
as high as 15--20\% (Bowyer et al.~1996). 
The produced light curve with
20 phase bins yields the pulsed fraction $\fp =27\pm 5$\%.

\begin{figure}
\epsfxsize=8.5cm
\epsffile[0 70 400 400]{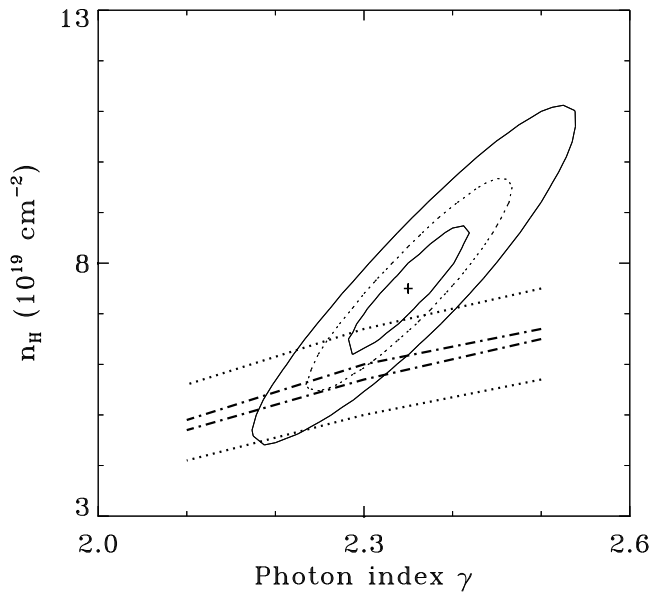}
\caption[ ]{
68\%, 90\% and 99\% confidence contours for the power-law fit to
the PSPC spectrum combined from two data sets.
Thick dot-dashes are the lines of minimum and maximum DSI fluxes
if only statistical errors are taken into account;
thick dots correspond to possible $\pm 15\%$ systematic
uncertainty of the DSI effective area.
}
\label{fig1}
\end{figure}

\subsection{Power-law fit}
A power law  with the photon index
$\gamma=2.35\pm 0.20$, intrinsic source luminosity 
$L_x=(7.3\pm 0.8)\times 10^{30}\, d_{180}^2$~erg~s$^{-1}$ in 0.1-2.0~keV 
and $n_H=(7.5\pm 2.3)\times 10^{19}$~cm$^{-2}$ provides an
acceptable ($\chi_\nu=1.08$)
fit to the combined PSPC spectrum. 
The parameter values are generally consistent with 
those obtained by BT93 and HMM96: $\gamma=2.6\pm 0.2$,
$n_H=(1.4\pm 0.5)\times 10^{20}$~cm$^{-2}$, $L_x=
9.9\times 10^{30}\, d_{180}^2$~erg~s$^{-1}$, and
$\gamma=2.45\pm 0.25$, $n_H=(9.2\pm 4.8)\times 10^{19}$~cm$^{-2}$,
$L_x=8.3\times 10^{30}\, d_{180}^2$~erg~s$^{-1}$, respectively.
The best power-law fit to the combined PSPC spectrum yields the HRI count
rate $0.044$~s$^{-1}$, which is compatible with the measured value
of $0.043\pm 0.004$~s$^{-1}$.

Folding the same model with the $EUVE$ DSI response
gives the count rate $0.00747$~s$^{-1}$, 
which is about 20\% lower than the observed value. Following HMM96, we  
tested whether or not the single power-law fits to the PSPC spectrum
can be consistent with the DSI flux. 
We folded each of the trial spectra on a grid of the 
fitting parameters $\gamma$ and $n_H$ 
through the effective area curve of DSI and obtained a corresponding
grid of predicted DSI count rates. Figure~1 presents the 68\%, 90\%  
and 99\% confidence contours of the PSPC fits in the $\gamma$-$n_H$ plane, 
together with the bands allowed by estimated 
systematic uncertainties of the DSI response.
Figure~1 shows that our joint PSPC+DSI analysis restricts 
the hydrogen column density to
$n_H=(5.4-7.1)\times 10^{19}$~cm$^{-2}$,
at a 90\% confidence level. 
This restriction is compatible with that of HMM96,
$n_H=(5-8)\times 10^{19}$~cm$^{-2}$, obtained with almost thrice
lower number of the PSPC counts.
It should be noted, however, that if one adopts
the 2\% statistical uncertainty for the DSI domain, 
it does not overlap the 68\% PSPC confidence ellipse,
so that the joint analysis of the PSPC and DSI data 
hints that the power-law spectral model may be only {\em marginally}
acceptable. 
This can be considered as one more reason
to test {\em thermal} models for the soft X-ray radiation
of \psr. 

\begin{figure*}
\epsfxsize=16cm
\epsffile[10 0 800 450]{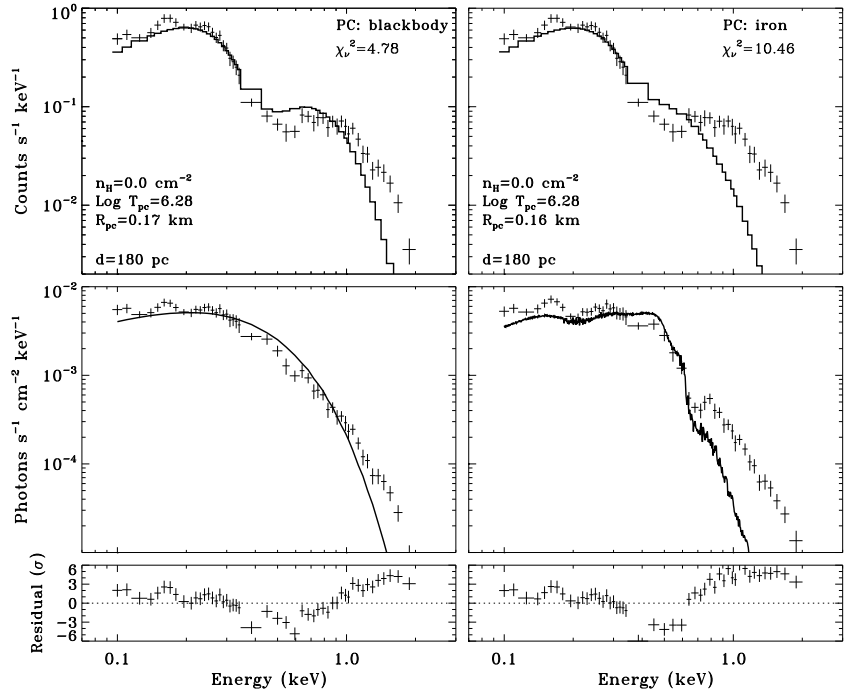}
\caption[ ]{
Single PC model fits (blackbody radiation and iron
atmosphere) to the PSPC count rate spectrum,
with values of best-fit parameters.}
\end{figure*}

\section{Single-temperature polar cap model}

In this Section we analyze the above-described data in terms of
the simplest one-component thermal model. We assume that
the observed radiation is emitted by two symmetric, uniformly heated PCs
around the (dipole) magnetic poles. We consider three chemical
compositions for the PC atmospheres, and, for the sake of comparison,
the blackbody model for the PC radiation. The fitting parameters for
this model are $n_H$, the PC (effective) temperature $\Tpc$, and 
the ratio $\Rpc /d$. The radius $\Rpc$ is defined as the 
distance from the cap edge to the magnetic axis.
We fix the distance to the pulsar, $d=180$~pc
(a most probable value, according to Sandhu et al.~1997),
and present the results for $\Rpc$ and PC luminosity for
this distance. Since the atmosphere spectra and, particularly,
light curves depend on the viewing angle, the directions
of the magnetic and rotation axes are to be specified.
We adopt, for most of our fits,
$\zeta =40^\circ$ for the angle between the rotation axis and 
the line of sight, and $\alpha=35^\circ$ for the angle  between the rotation
and magnetic axes (Manchester \& Johnston 1995).
To specify the gravitational acceleration for
calculating the atmosphere models, and the NS mass-to-radius
ratio for account of the gravitational bending of the
photon trajectories, we adopt $\mns =1.4 M_\odot$
and $\rns =10$~km for the NS mass and radius.
All temperatures and radii here and below are as measured at the
pulsar surface.

\subsection{PSPC spectral fits}
Fitting the combined PSPC spectrum with four models for
the PC radiation yields the following results.

{\em Blackbody}.--- 
In agreement with what was found by BT93 and confirmed later
by HMM96 for the first PSPC data set, the single blackbody model fails to
fit the combined PSPC spectrum (minimum $\chi_\nu^2 = 4.78$) because 
the blackbody spectrum is too soft (steep) at high energies and 
leaves too large residuals at $E\gapr 0.9$~keV (Fig.~2).
Formally, the fit yields $n_H=(0.0\pm 1.1)\times 10^{19}$ cm$^{-2}$, 
$\log\Tpc =6.28\pm 0.03$ and $\Rpc =(0.17\pm 0.02)\, d_{180}$~km.
This gives the bolometric luminosity of two PCs 
$L_{\rm bol}=(1.36\pm 0.25)\times 10^{30}\, d_{180}^2$~erg~s$^{-1}$.

{\em Iron atmosphere}.---
Because of numerous spectral lines
quasi-randomly distrubuted at $\sim 0.3-3.0$~keV, 
the smoothed spectra emitted by iron atmospheres
in the soft X-ray range are not much different from
blackbody spectra at the same temperatures
(Rajagopal \& Romani 1996; ZPS96). The spectral 
lines of iron atmosphere spectra 
cannot be detected by $ROSAT$ due to its poor energy resolution.
The spectral fit with the iron atmosphere model (right panel of Fig.~2) is 
even worse than with the blackbody: minimum $\chi_\nu^2=10.46$.  
The lower quality of this fit is caused by the fact (ZPS96) 
that the iron atmosphere spectra are softer the blackbody spectra
in the soft X-ray range at temperatures $\sim 10^6$~K.   

{\em Hydrogen atmosphere}.--- 
In comparison with the two cases above, the PC model with hydrogen
composition gives a much better fit, with minimum
$\chi_\nu^2=1.68$ (Fig.~3). Since the hydrogen
atmosphere spectra, as well as any other thermal spectra, 
have an intrinsic decrease 
towards low energies, they fit the observed spectrum at
much lower interstellar hydrogen density, 
$n_H=(0.0\pm 1.7)\times 10^{19}$ cm$^{-2}$.
The other parameters are $\log\Tpc =5.94\pm 0.03$ and
$\Rpc =(0.77\pm 0.13)\, d_{180}$~km. 
Since the high-energy tail of the radiation emitted 
from hydrogen atmospheres is harder than the blackbody tail, 
the resulting PC temperature is 2.2 times lower
than the temperature of the blackbody fit.
To provide the same flux, $\Rpc$ becomes 4.5 times greater
approaching the conventional estimate of $\sim 1.9$~km.
The corresponding bolometric luminosity of two PCs is
$L_{\rm bol}=(1.22\pm 0.22)\times 10^{30}\, d_{180}^2$~erg~s$^{-1}$, 
comparable with that for the blackbody model.
We checked that the fitting parameters are very stable with respect to
the choice of energy channels (binning) of the PSPC count rate spectrum
--- only the minimum value of $\chi_\nu^2$ varies slightly.
For example, the fit in 20-140 channels ($0.2-1.4$~keV range) with 19 bins
yields the same values of $n_H$, $\Tpc$ and $\Rpc$,
with minimum $\chi_\nu^2=1.16$.

\begin{figure*}
\epsfxsize=16cm
\epsffile[10 0 800 450]{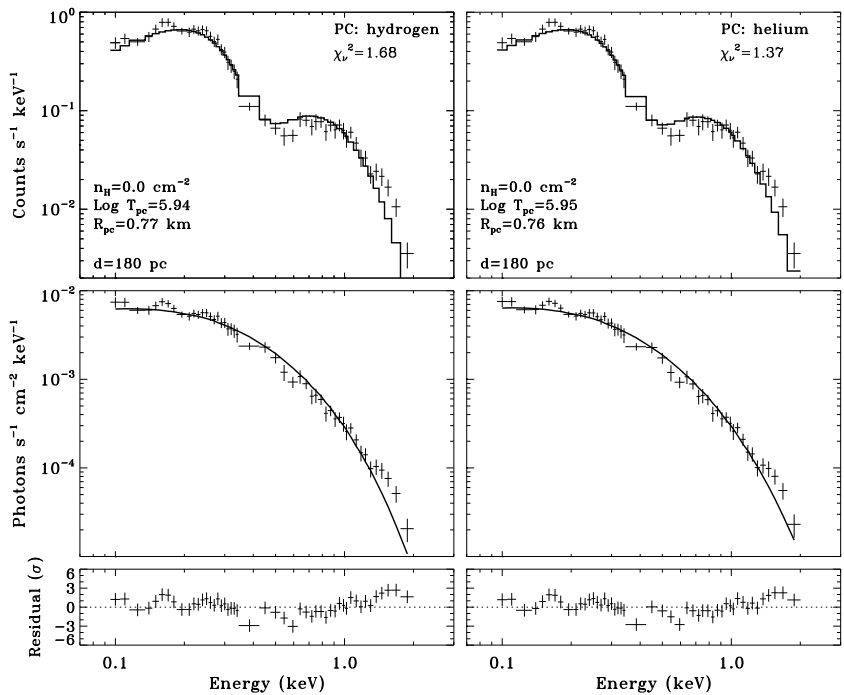}
\caption[ ]{
The same as in Fig.~2 for the hydrogen and helium atmospheres.}
\end{figure*}

{\em Helium atmosphere}.---
For helium-covered PCs (right panel of Fig.~3), the quality of the fit
and the values of the model parameters are similar to those
for the hydrogen composition.
A smaller value of the minimum $\chi_\nu^2=1.37$ is due to
the fact that at $T\sim 10^6$~K the spectrum of 
the helium atmosphere is slightly
harder than that of the hydrogen atmosphere (ZPS96). Generally,
the fits with hydrogen and helium models are almost indiscernible. 

Although the hydrogen and helium PC models fit the PSPC spectrum much 
better than the blackbody model, there still is an excess
of the observed radiation at $E\gapr 1.5$~keV (see Fig.~3).
The excess can be partly due to 
systematic errors of the high-energy PSPC response
which may be as high as 20\%. Irrespective of that,
the excess can be naturally explained by simplicity of the
single-temperature PC model --- if there is a smaller
area of a greater temperature within the PC, it can easily
provide the excess high-energy quanta, the possibility we 
explore in Section~4.
The single-temperature approximation, however, enables us to
show that the spectrum can be fitted with a purely thermal
model provided that the PCs are covered by a hydrogen or
helium atmosphere, whereas the hypothesis about the iron surface
of the PCs can be firmly rejected. (A similar conclusion
has been drawn by Rajagopal \& Romani (1996) who, however,
obtained substantially different PC parameters because
they fitted the observed spectrum without allowance
for the limb-darkening and light bending.)
We do not expect that 
other heavy-element compositions of the NS surface could
fit the observed spectrum --- they should also lead to
emitted spectra much softer that the spectra of hydrogen or 
helium atmospheres. We consider the hydrogen atmosphere
more natural that the helium one (e.~g., if the atmosphere
was formed due to accretion of the hydrogen-rich interstellar
matter or matter from the binary companion), and we 
explore consistency of this hypothesis with the HRI and DSI
results, as well as with the light curves observed with 
the three instruments.

\begin{figure*}
\epsfxsize=16cm
\epsffile[10 225 800 450]{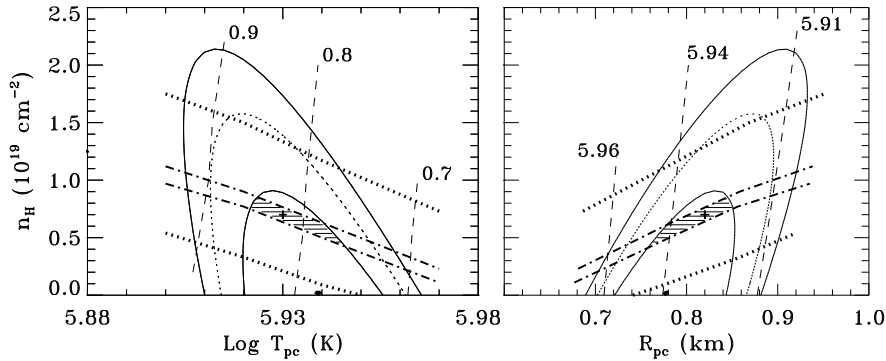}
\caption[ ]{
Projections of the 68, 90 and 99\% confidence domains
onto two parameter plains for the hydrogen PC fit to
the PSPC spectrum. Thin dashes correspond to constant values
(numbers near the lines) of third parameter.
Filled squares show the best-PSPC-fit parameters (cf.~Fig.~3).
Thick dot-dashes and dots are the lines of constant
DSI fluxes (cf.~Fig.~1).
The shaded regions give the model parameters allowed by the count rates
of all the three instruments. Crosses correspond to the ``best PC model''.
}
\end{figure*}

\subsection{Analysis of the $ROSAT$ HRI and $EUVE$ DSI count rates}
Figure~4 shows the confidence contours 
for the parameters of the hydrogen PC fit to the PSPC spectrum. 
We checked that any point from the 99\% PSPC 
confidence region produces the model HRI count rate within the observed range.
The best parameters of the PSPC spectral fit give a too
high count rate for the $EUVE$ DSI, 0.01136~s$^{-1}$, which 
does not even get into the $\pm 15$\% systematic 
uncertainty range around the observed value. 
The reason is that the formal best fit yields $n_H=0$ (no interstellar
absorption), which clearly is not a realistic value.
However, $n_H$ is determined with some uncertainties,
and even a small value of $n_H$ can noticeably decrease
the DSI flux and make the DSI and PSPC count rates consistent with each other.
Similar to the case of the power-law model (Section~2), we computed the bands 
in the parameter planes allowed by the observed DSI flux.
The thick dot-dash lines in Fig.~4 correspond to the
DSI count rates of $0.00973\pm 0.00017$~cnts/s. The shaded regions show
the domains of the model parameters 
allowed by all the three instruments at the 68\% confidence level. 
If we adopt the $\pm 15\%$
uncertainty for the observed DSI count rate (thick dots), the allowed
regions become much broader. 

Thus, there exist a domain of the model parameters
compatible with all of the detected count rates and with the PSPC spectra:
$n_H=(0.5-0.9)\times 10^{19}~{\rm cm}^{-2}$, 
$\log\Tpc =5.92-5.95$ and  
$\Rpc =(0.75-0.85)\, d_{180}~{\rm km}$. 
The corresponding
area and bolometric luminosity of two PCs are $A=(3.5-4.5)\, d_{180}^2$~km$^2$
and $L_{\rm bol}=(1.2-1.3)\times 10^{30}\, d_{180}^2$~erg~s$^{-1}$.
Similar to the case of the power-law fit, 
the main role of the $EUVE$ data is that these data enable us
to restrict the $n_H$ value; the difference is that for the thermal
PC model the range of jointly acceptable $n_H$ values lies
above the best-PSPC-fit value, and they are an order of magnitude
lower than for the joint power-law fit. This implies 
a higher ionized fraction, and a lower mean number density
of the interstellar hydrogen in the direction to \psr~ (see Discussion).
The center of the allowed three-dimensional parameter domain 
is given, approximately, by the following values:
$n_H=0.7\times 10^{19}$~cm$^{-2}$, $\log\Tpc =5.93$ and
$\Rpc =0.82\, d_{180}$~km, which can be considered
as a best joint PSPS+HRI+DSI fit (marked with crosses in Fig.~4).
For the sake of brevity, we call this set of parameters
``{\em the best PC model}'' for the rest of this Section.

\begin{figure}
\epsfxsize=8.5cm
\epsffile[0 100 400 800]{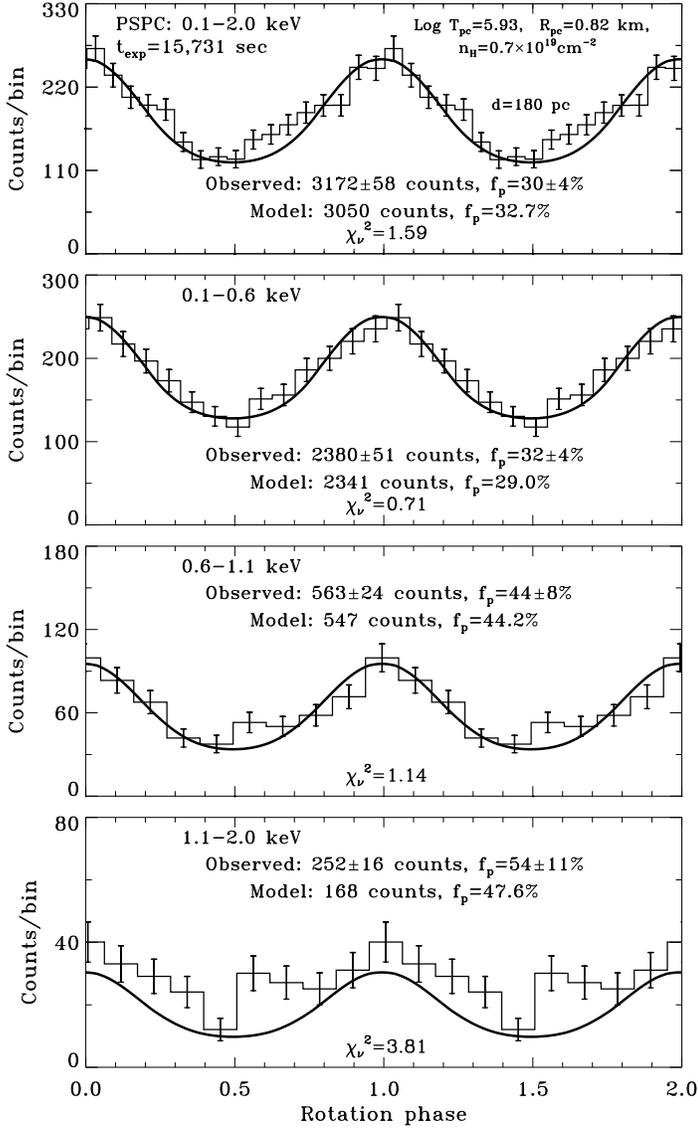}
\caption[ ]{
Observed and model PSPC light curves in four energy ranges.
The model curves correspond to the ``best PC model''
(model parameters in the upper panel).
}
\end{figure}

\subsection{Light curve analysis}
Now we should test whether the PC model obtained from
the analysis of the PSPC spectrum and HRI and DSI count rates
fits the light curves detected with these instruments.
To do that, we computed the spectra for 80
rotation phases, $0\leq \phi \leq 1$, for the best PC model and
convolved the spectra at each phase
with the PSPC, HRI and DSI responses. Then, we binned these 
model count rates into the 
the phase bins adopted for the observed light curves to
compare the light curves
with the aid of the $\chi^2$-test. The $\chi^2$ value depends
on how we co-align the phases for the model and the observed
light curves. Since the reference phases of the observed light curves
were chosen arbitrarily, we may vary the 
phase shift between the model and observed light curves
to minimize $\chi^2$. (Of course, the shifts in different
energy ranges should not differ from each other by more than
a fraction of the bin width; ideally, the difference should
be comparable with, or less than, the instrument time resolution).

\begin{figure}
\epsfxsize=8.5cm
\epsffile[0 50 400 400]{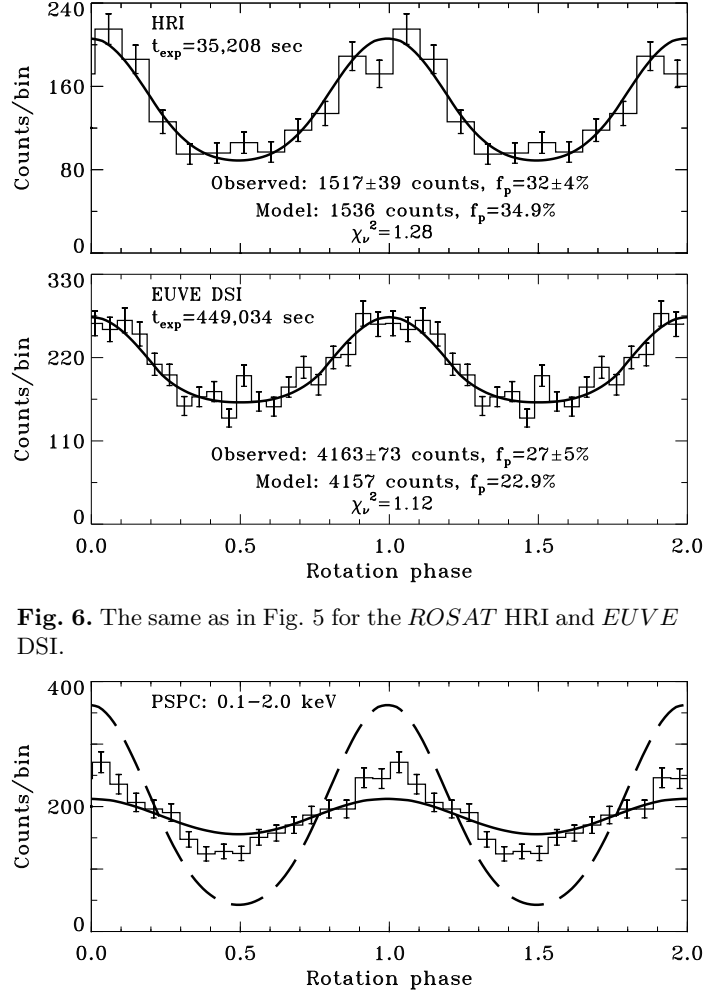}
\caption[ ]{
The same as in Fig.~5 for the $ROSAT$ HRI and $EUVE$ DSI.
}
\end{figure}

\begin{figure}
\epsfxsize=8.5cm
\epsffile[0 25 400 200]{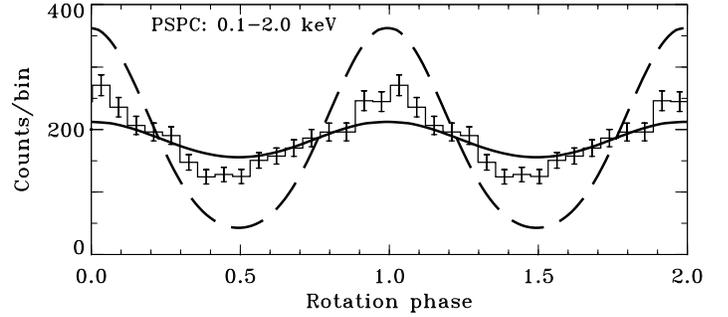}
\caption[ ]{
Two examples of hydrogen PC light curves:
for another set of the angles ($\zeta=24^\circ$ and $\alpha=20^\circ$; solid)
and for the best PC model without allowance for the
gravitational bending (dashes).
}
\end{figure}

Figure~5 presents the adjusted model and observed PSPC light curves for the
four energy ranges. We define the zero rotation phase 
as the phase when the angle $\theta$ between 
the magnetic axis and the line of sight
equals its minimum value, $\theta_{\rm min}=\zeta-\alpha =5^\circ$;
this phase corresponds to the maximum of the
model light curve, whereas the minimum lies at $\phi=0.5$,
corresponding to $\theta=\theta_{\rm max}=\zeta+\alpha=75^\circ$.
The difference of the phase shifts applied to minimize $\chi^2$
does not exceed 1/80 of the rotational period, which  corresponds to 0.07~ms,
about one half of the the $ROSAT$ time resolution.
As is seen in Fig.~5, the best PC model yields the 
light curves in excellent agreement with those observed
in the $0.1-0.6$ and $0.6-1.1$~keV ranges ---
not only the pulsed fractions are within the statistical
uncertainties, but also the $\chi_\nu^2$ values are small
enough. For the $1.1-2.0$~keV range,
the model $\fp$ is within the observed uncertainty, but the fit is 
formally unacceptable ($\chi_\nu^2=3.81$).
This is not surprising in view of the aforementioned deficit of PC counts
with respect to observed in this energy range. 

Figure~6 shows the observed $ROSAT$ HRI
and $EUVE$ DSI light curves 
It is worth noting that the difference between the phase
shift applied to adjust the observed PSPC and HRI light
curves to the model ones is as small as 1/80.
The coincidence of the model and observed light curves
is fairly good for both HRI and DSI.

The results above were obtained for $\zeta=40^\circ$, $\alpha=35^\circ$,
and the standard NS mass and radius. The angles, however,
are not very certain as this pulsar
shows a complicated variation of linear polarization angle
across the eight-component mean radio pulse (Manchester \& Johnston 1995). 
In principle, the interpretation of the soft X-ray data adopted
in the present paper can reduce this uncertainty --- 
one can fit the data with the PC model
considering $\alpha$ and $\zeta$ as fitting parameters.
These angles, as well as the (unknown) ratio 
$\mns/\rns$, affect insignificantly the spectral fit parameters
($n_H$, $\Tpc$ and $\Rpc$). 
However, the shape of the light curves and $\fp$ are very sensitive
to the choice of $\alpha$, $\zeta$ and $\mns/\rns$. 
For example, an alternative
set of the angles, $\zeta=24^\circ$ and $\alpha=20^\circ$,
obtained with another treatment of polarization data 
(Gil \& Krawczyk 1997), results, for $\mns/\rns=0.14~M_\odot/{\rm km}$,
in a too low pulsed fraction, $\fp=15.3\%$ in the whole PSPC energy range,
so that the PC model is in clear disagreement with observations
($\chi_\nu^2=3.52$) --- see Fig.~7.
To provide an acceptable fit at this set of angles,
the pulsar mass is to be
very low, $\mns < 0.5 M_\odot$ at any $\rns$ allowed by the equations of 
state of NS interiors (Pavlov et al.~1997).
One more example shown in Fig.~7 demonstrates the importance of 
the gravitational bending of photon trajectories: 
if this effect is neglected, the best PC model yields a
light curve with strong pulsations, $\fp=76.8$\%, resulting in
a great deviation,  $\chi_\nu^2=29.1$.
The mass-to-radius ratio can be also considered as
a fitting parameter to provide  better coincidence 
of the light curves and put constraints
on the equation of state of the NS interiors.

\begin{figure*}
\epsfxsize=16cm
\epsffile[10 50 800 450]{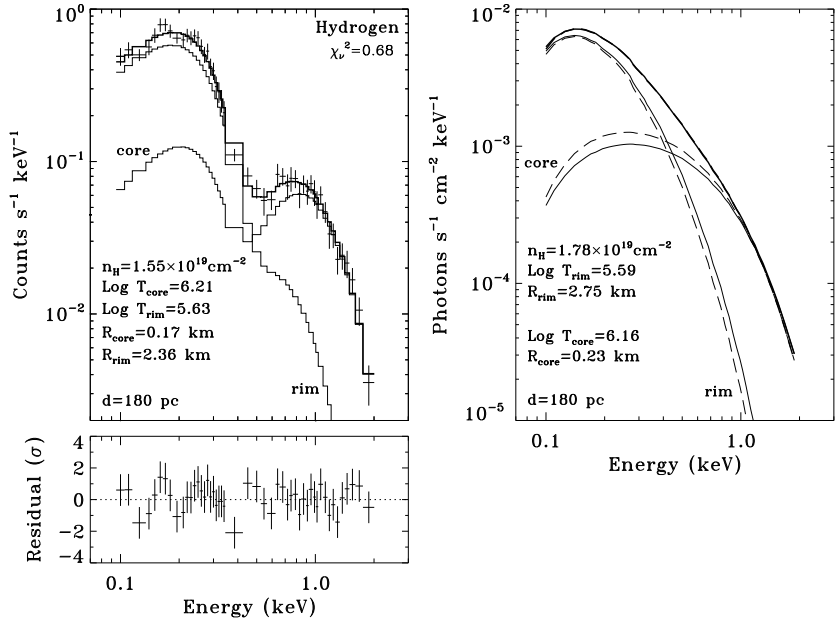}
\caption[ ]{
Two-component (``core+rim'') fit to the PSPC spectrum
with five free parameters (best values in the upper-left panel).
Photon spectral fluxes are shown in the right panel for the
five-parameter fit (solid curves) and for the fit with fixed
values of $\Tc$ and $\Rc$ (dashes).
}
\end{figure*}

\section{Two-temperature polar cap model}

Motivated by the results of Section~3, we 
examine a more complicated PC model which allows, in the
first approximation, for a non-uniform temperature distribution
along the PC surface.  We adopt a two-step approximation, ``core+rim'',
with $\Tc >\Tr$ and $\Rc <\Rr$.
The definitions of $\Rc$ and $\Rr$ are analogous to $\Rpc$ in
Section~3. This model contains, together with $n_H$, five fitting parameters. 
It is natural to expect that $\Tr < \Tpc < \Tc$ and
$\Rc < \Rpc < \Rr$, where $\Tpc$ and $\Rpc$ are the
parameters of the one-component PC model, so that the rim and the core 
should be responsible for the soft and hard spectral
components of this two-component model.
We retain the same assumptions on the pulsar angles, mass, radius
and distance as earlier.

\subsection{Iron atmosphere}
As one could expect for a two-component model, the iron atmosphere
model yields an acceptable fit of the PSPC spectrum 
(minimum $\chi_\nu^2=0.72$, $\nu=44$), with the  best-fit
parameters $n_H=7.92\times 10^{19}$~cm$^{-2}$, $\log\Tc =
6.57$, $\log\Tr =5.67$, $\Rc =0.04\, d_{180}$~km 
and $\Rr =8.36\, d_{180}$~km. However, the value of $\Rr$
is very close to the adopted pulsar radius $\rns =10$~km. This means
that the soft component is emitted from almost the {\em whole}
NS surface with the uniform temperature $\Tr$. 
By virtue of this, {\em the rim cannot 
produce the pulsed fraction} $\fp\simeq
30$\% observed in the 0.1--0.6~keV range
(the model $\fp$ does not exceed a few percent). 
Since the light curves cannot be fitted with the same model
as the spectra, we conclude that the hypothesis about the iron
NS surface is incompatible with the thermal model for the
soft X-ray radiation of \psr.
Thus, we exclude the iron atmosphere model
from further consideration, as well as the two-component
blackbody model which also leads to $\Rr\simeq \rns$.

\begin{figure}
\epsfxsize=8.5cm
\epsffile[0 70 400 400]{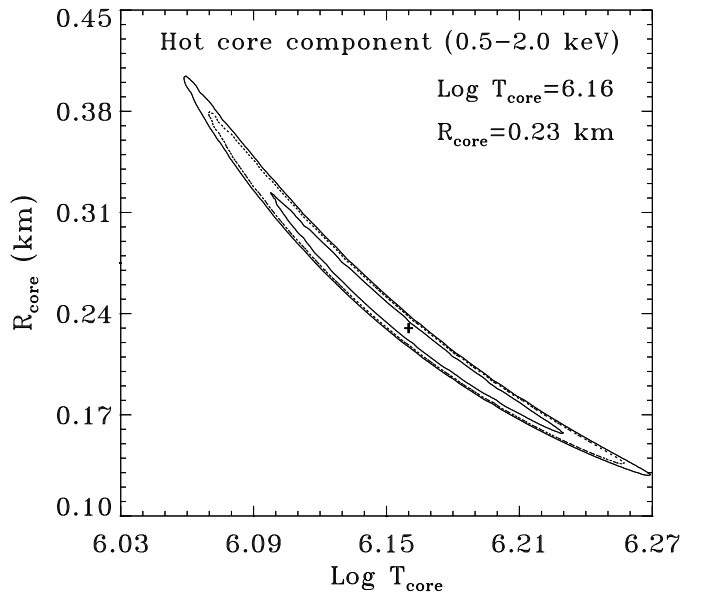}
\caption[ ]{
Confidence contours for the parameters of
hard (core) component in the two-component PSPC spectral fit.
}
\end{figure}

\subsection{Hydrogen atmosphere}
The model count rate spectra in Fig.~8 
correspond to the best spectral fit with the following
parameters: $n_H=1.55\times 10^{19}$~cm$^{-2}$, $\log\Tc =6.21$, 
$\log\Tr =5.63$, $\Rc =0.17\, d_{180}$~km and  
$\Rr =2.36\, d_{180}$~km  (minimum $\chi_\nu^2=0.68$, $\nu=44$).
The corresponding bolometric luminosity of two rims and cores, 
$L_{\rm bol}= 1.37\times 10^{30}\, d_{180}^2$~erg~s$^{-1}$ (51\% and 49\% from
the cores and rims, respectively), 
is close to that of the single-temperature PC model. 

\begin{figure*}
\epsfxsize=16cm
\epsffile[10 225 800 450]{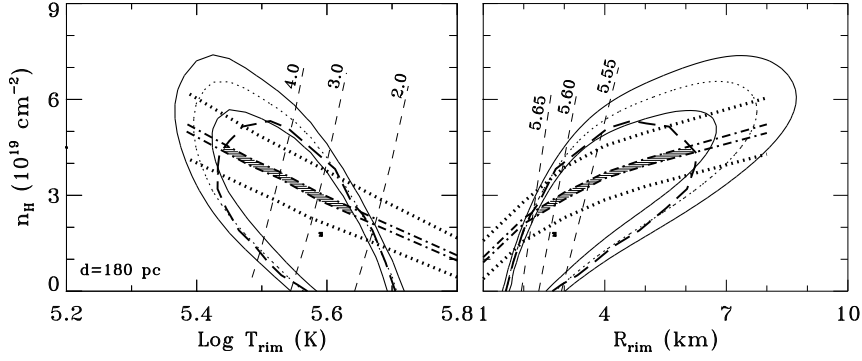}
\caption[ ]{
Projections of 68, 90 and 99\% confidence domains for
the three-parameter PSPC spectral fit
of the soft (rim) component (thin solid and dot lines),
with fixed best core parameters (cf.~Fig.~9).
Thick dashed lines show restriction (68\% confidence
level) obtained from the light curves in the whole PSPC
energy range. Other notations are the same as in Fig.~4.
}
\end{figure*}

\begin{figure}
\epsfxsize=8.5cm
\epsffile[0 100 400 800]{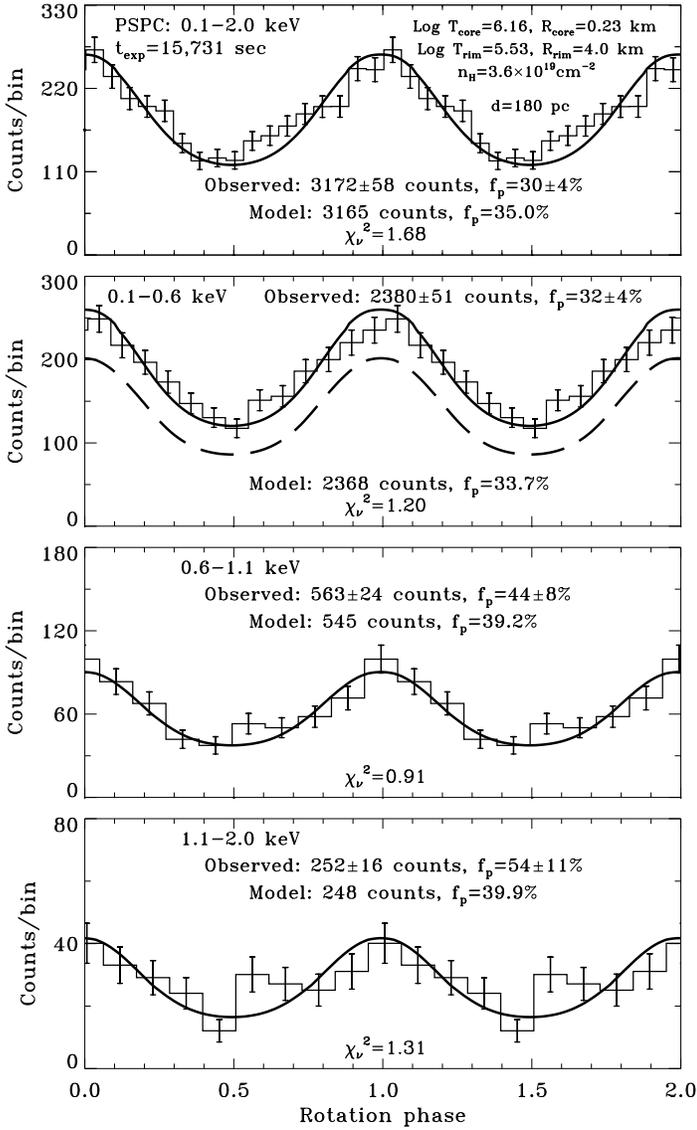}
\caption[ ]{
The same as in Fig.~5 for the ``core+rim'' two-component fit.
The dashes show the the contribution to the o.1--0.6 keV
light curve from the rim component
}
\end{figure}

However, in spite of quite realistic values of these 
parameters, they are very uncertain, which follows from 
large formal errors provided by the fitting code.
Although these errors,
calculated from the derivatives of the fit statistics,
are not reliable, they indicate
that the parameters are strongly correlated,
and the best fit gives only qualitative estimates of the parameter values.
In principle, the parameter ranges can be reduced with 
the use of the $ROSAT$ HRI and
$EUVE$ DSI data, but it would require an extremely
time-consuming analysis of five-dimensional confidence
volumes. A practical way to solve the problem is prompted by the shapes
of the best-fit component spectra --- we see that
the hot core component gives the main contribution 
at higher energies, above 0.4--0.6~keV, whereas
the rims are responsible for the flux at lower energies.
Therefore, we can first fit the hard channels separately
with a single-temperature cap model 
(notice that $n_H$ of expected magnitudes does not affect
spectra at these energies), and then fix some set(s) of $\Tc$ and $\Rc$
and fit the rim component with the three fitting parameters,
$n_H$, $\Tr$ and $\Rr$.

The first step with 21 energy bins in the 53--200 channels
gives the best-fit parameters of the hot core, $\log\Tc =6.16\pm 0.07$ and
$\Rc=(0.23\pm 0.09)\, d_{180}$~km ($\chi_\nu^2=0.92$, $\nu=19$), 
in accordance with the two-component 
results. Figure~9 demonstrates the confidence 
contours for this fit.

At the second step, we fixed the best pair of $\Tc$ and $\Rc$
obtained at the first step and 
fitted the total PSPC count rate spectrum with the
``core+rim'' model (the best-fit parameters 
are given in the right panel of Fig.~8).
We see that the separate fitting of the hard and soft (core and rim)
components yields spectra fairly close to those obtained
with the five-parameter fit. 

The projections of the three-dimensional confidence regions for
the parameters of the soft component are shown in Fig.~10.
The regions are still too broad, in spite of
the reduced number of the fitting parameters, because the
parameters are strongly correlated (e.~g, $n_H\propto \ln\Rr$ at
a given $\Tr$) due to narrowness of the soft-energy range. In particular, 
the 99\% confidence surface allows the outer rim radius
to be comparable with $\rns$ at lower $\Tr$ and higher $n_H$.
However, 
large values of $\Rr$ are not consistent with the strong time modulation of
the observed flux at low energies, and upper limits for $\Rr$
and $n_H$ (and a lower limit on $\Tr$) are to be obtained from
the light curve analysis. 
Since the values of $\Rr$ involved in the light curve fitting
are relatively large, we have to integrate over the rim surface
making use of the exact equations for the gravitational bending
(Zavlin et al.~1995), instead of previously used approximate
approach for $\Rr \ll \rns$.
Fitting the PSPC light curve in the whole energy range with the
two-component model restricts the radius: $\Rr < 6.2$, 6.7 and 7.1~km;
temperature: $\log\Tr > 5.45$, 5.42  and 5.39;
and hydrogen column density: $n_H/(10^{19}~{\rm cm}^{-2}) < 4.4$, 4.8
and 5.2; at 68\%, 90\% and 99\% confidence levels, respectively.
The projections of the 68\% confidence region are shown in Fig.~10.

Further constraints on the fitting parameters follow from
the comparison with the $EUVE$ DSI count rate (HRI
count rate is consistent with the PSPC fits and does not
provide additional restrictions of the fitting parameters).
The DSI count rate corresponding to the best PSPC fit (with
the hard and soft components fitted separately) is
0.01196~s$^{-1}$, i.~e., 23\% higher than the observed value. 
We used the same approach as for the single-temperature
PC model and found the three-dimentional volumes in the parameter
space compatible with the measured DSI count rates with the 2\%
(statistical) and 15\% (systematic) uncertainties (thick
dash-dot and dot curves in Fig.~10). 
The shaded areas in Fig.~10 delimit projections of the
parameter volume compatible with the data collected with
all the three instruments at the 68\% confidence level
(neglecting poorly-known systematic uncertainties).
We chose a point with $n_H=3.6\times 10^{19}$~cm$^{-3}$,
$\log\Tr=5.53$ and $\Rr =4.0\, d_{180}$~km (``{\em best core+rim model}''). 
The bolometric luminosity for this model is
$L_{\rm bol}=1.57\times 10^{30} d_{180}^2$~erg~s$^{-1}$.

Following the approach used for the one-component PC model, we plotted the 
two-component light curves corresponding to this model
(Figs.~11 and 12). The comparison
shows a satisfactory agreement with the observed light curves.
We see that the pulsations at lower energies are stronger
for the ``core+rim'' model than for the single-temperature PC model 
($\fp=33.7$\% vs. 29.0\% for the 0.1-0.6~keV PSPC range, 
and 25.7\% vs. 22.9\% for DSI).
This is due to the fact that the rim, 
which mainly contributes at $E\lapr 0.5$~keV (Fig.~8), has
a lower temperature, $\Tr <\Tpc$, resulting in
higher anisotropy of the emergent radiation (ZPS96). 
The rim emits 74\% of all the PSPC counts with $\fp=37.1$\%
at $E=0.1-0.6$~keV, and 87\% of counts with $\fp=31.6$\% in the DS range. 
The soft component gives less than a half of the total HRI counts, 
and we obtain almost the same light curve fits for the 
single-temperature PC and ``core+rim'' models (cf.~Figs.~6 and 12).
In the whole PSPC range there is a bigger share of the total radiation 
with $\fp=36.4$\% coming from the rim,
that leads to higher pulsations in the summed light curve
than for the simplest PC model ($\fp=35.0$\% vs. 32.7\%, respectively).

\begin{figure}
\epsfxsize=8.5cm
\epsffile[0 50 400 400]{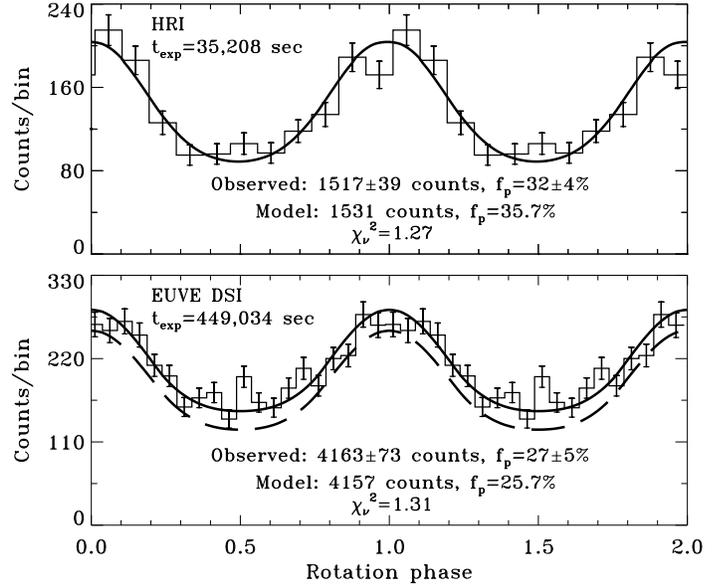}
\caption[ ]{
The same as in Fig.~11 for the $ROSAT$ HRI and $EUVE$ DSI,
with separate contribution from the rim component (dashes).
}
\end{figure}
Finally, we can conclude that 
the two-component hydrogen ``core+rim'' model 
satisfactorily matches to the spectral data and the light curves 
provided by all the three instruments.

\subsection{Helium atmosphere}
As expected, this model yields virtually the same results as
the hydrogen atmosphere model. For instance,
the best five-parameter fit gives $n_H=1.98\times 10^{19}$~cm$^{-2}$,
$\log\Tc =6.14$, $\log\Tr =5.57$, $\Rc =0.26\, d_{180}$~km,
and $\Rr =2.94\, d_{180}$~km, with $\chi_\nu^2=0.70$.
Since, in addition, the hydrogen composition 
looks more natural, the helium atmosphere model
does not require a detailed separate analysis.

\section{Discussion}

\subsection{Comparison with predictions of radio pulsar models}
It is interesting to compare the inferred PC properties with
those predicted by different radio pulsar models.
The {\em slot-gap model} (Arons \& Scharlemann 1979; Arons 1981)
predicts the following energy flux of relativistic particles
incident on the stellar surface
(Arons 1981)
\begin{eqnarray}
F\simeq 1\times 10^{19}  \left(\frac{\mu_{26}}{3.4}\right)
\left(\frac{5.75~{\rm ms}}{P}\right)^{19/8} 
\left(\frac{4}{f_\rho}\right)^{1/4} \\
\left(\frac{10~{\rm km}}{\rns}\right)^{9/8} 
\left(\frac{\sin\alpha}{\sin 35^\circ}\right)^{5/4}~{\rm erg~cm^{-2}~s^{-1}}~,
\nonumber
\end{eqnarray}
where $\mu_{26}$ is the magnetic moment in units of $10^{26}$~G~cm$^3$,
$f_\rho$ is the the ratio of the dipole radius of curvature to the actual
radius of curvature, $1\lapr f_\rho \lapr 7$ for the parameters
of \psr. The model is applicable for $P < 3.1 R_{10}^{-3/17}
f_\rho^{10/17}$~ms; this corresponds to $f_\rho > 2.85
R_{10}^{0.3}$ for $P=5.75$~ms, and we adopt an intermediate value $f_\rho=4$
for estimates. If most of this energy is reradiated, the effective
temperature and luminosity are
\begin{eqnarray}
T_{\rm eff}=6.5\times 10^5 \left(\frac{\mu_{26}}{3.4}\right)^{1/4}
\left(\frac{5.75~{\rm ms}}{P}\right)^{19/32}
\left(\frac{4}{f_\rho}\right)^{1/16} \\
\left(\frac{10~{\rm km}}{\rns}\right)^{9/32} 
\left(\frac{\sin\alpha}{\sin 35^\circ}\right)^{5/16}~{\rm K}~,
\nonumber
\end{eqnarray}
\begin{eqnarray}
L_{\rm cap}=0.6\times 10^{30} \left(\frac{f_a}{0.5}\right)
\left(\frac{\mu_{26}}{3.4}\right)
\left(\frac{5.75~{\rm ms}}{P}\right)^{27/8} \\
\left(\frac{4}{f_\rho}\right)^{1/4} 
\left(\frac{\rns}{10~{\rm km}}\right)^{15/8} 
\left(\frac{\sin\alpha}{\sin 35^\circ}\right)^{5/4}~{\rm erg~s^{-1}},
\nonumber
\end{eqnarray}
where $f_a$
is the ratio of the polar cap area 
to its canonical value, $11.4 R_{10}^3 (5.75~{\rm ms}/P)$~km$^2$
($f_a\simeq 0.5$ for the original slot-gap model which implies
that the particle acceleration zone is associated with
``favorably curved'' open field lines).
We see that the predicted temperature is $20-30\%$ lower than the
temperature $\Tpc$ inferred from the single-temperature PC model,
but it is higher than $\Tr$. The predicted luminosity is in
excellent agreement,
for $f_\rho=4$ and $f_a=0.5$, with the observed value.
Since the observed luminosity is almost independent of the 
PC temperature distribution
($2L=(1.0-1.5)\times 10^{30}$~erg~s$^{-1}$ for both the single-temperature
and two-temperature PC models), it is the most suitable parameter
to compare with, and we should conclude that the slot-gap model agrees
very well with the suggested interpretation of the soft X-ray
radiation from \psr.

According to the {\em outer-gap model} (e.~g., Cheng, Ho \& Ruderman
1986), \psr~ is near the pulsar ``death line'', $P_{\rm death}=
5.8 (B/3.4\times 10^8~{\rm G})^{5/12}$~ms, that implies  high efficiency
of the gamma-ray radiation produced by primary e$^\pm$ accelerated
in vacuum gaps in the outer magnetosphere of the pulsar.
The total (maximum) flux of relativistic e$^+$ (or e$^-$) 
impinging on the
PC from the starward end of the acceleration zone can be
estimated as
\be
\dot{N}\sim \frac{\mu \Omega^2}{2ec}=1.4\times 10^{31}
\left(\frac{\mu_{26}}{3.4}\right)
\left(\frac{5.75~{\rm ms}}{P}\right)^2~{\rm s}^{-1}~,
\ee
and the luminosity is $L_{\rm cap}=\dot{N}\gamma_f m_ec^2$,
where $\gamma_f$ is the Lorentz factor of particles when they reach
the NS surface. 
While traveling to the surface, the particles  loose
a fraction of their initial energy $\gamma_im_ec^2$
via curvature radiation, so that
(cf.~Halpern \& Ruderman 1993; HMM96)
\begin{eqnarray}
\gamma_f  = \left[\frac{1}{\gamma_i^3} +
\frac{2e^2\Omega}{m_ec^3}\, \ln \left(\frac{r_{\rm min}}{\rns}\right)
\right]^{-1/3} = \\ 
\left[\frac{1}{\gamma_i^3} +
\left(\frac{1}{3.65\times 10^6}\right)^3\left(\frac{5.65~{\rm ms}}{P}\right)
\ln \left(\frac{r_{\rm min}}{\rns}\right) \right]^{-1/3}~, 
\nonumber
\end{eqnarray}
where $r_{\rm min}$ is the minimum distance of the accelerator
from the NS center. Adopting $r_{\rm min} \simeq (1/3)r_{\rm lc}\simeq
90 (P/5.75~{\rm ms})$~km,
the first term in the
square brackets can be neglected if $\gamma_i\gg 3\times 10^6$
(the latter value is much lower than $\gamma_{\rm max}
\sim 10^7 - 10^8$ limited by the curvature radiation losses).
In this case, $\gamma_f\simeq 2.8\times 10^6$, and the expected PC luminosity,
$L_{\rm cap}\sim 3.2\times 10^{31}$~erg~s$^{-1}$,
exceeds the observed value by a factor of 40--60.
The model PC luminosity would agree
with the observed one only if we assume $\gamma_i\simeq \gamma_f\simeq
6\times 10^4 \sim 10^{-3}\gamma_{\rm max}$. 
Thus, we should conclude
that either the $e^\pm$ energies at the starward
accelerator end are much lower than those expected from
the curvature radiation losses, or there exists a mechanism reducing
the particle flux $\dot{N}$ which may reach the NS surface.
In particular, the estimate (5) assumes the relativistic
particles impinge onto the canonical PC area
$A=(\pi \Omega R^3/c)\simeq 11$~km$^2$. If the actual area is $40-60$
times smaller (a ``primary'' cap radius is $250-300$~m, compatible
with our $\Rc$), then, at the same (Goldreich--Julian) current density,
the estimated luminosity coincides with that observed.
An evidence of the reduced efficiency of the particle accelerator
is provided by the upper limit on the luminosity of
gamma-rays above 100~MeV, $L_\gamma < 1.7\times 10^{32} = 0.04
\dot{E}$ (Fierro et al.~1995).
Since the same particle accelerator is responsible for PC heating,
HMM96 suggest that an upper limit for the luminosity of one PC can be
evaluated as 0.04 times the above-estimated maximum value,
$L_{\rm cap}< 1.3\times 10^{30}$~erg~s$^{-1}$, very comfortable with
the observed range of $(0.6-0.8)\times 10^{30}$~erg~s$^{-1}$.

In the {\em model of Beskin, Gurevich and Istomin} (1993), 
the energy of relativistic particles impinging onto the PCs
per unit time is
\be
\dot{E}_{\rm surf}\simeq \frac{Q^2}{1+K}\dot{E}~,
\ee
where 
$Q=0.52 (P/5.75~{\rm ms})^{11/10}
(\dot{P}/1.99\times 10^{-20})^{-2/5}$ 
and $K$ is a multiplication coefficient,
the number of particles which are knocked out from the NS surface
by an impacting relativistic particle. The temperature of the
polar cap is estimated as
\be
T_{\rm eff}=\frac{3.8\times 10^6}{(1+K)^{1/4}}
 \left(\frac{P}{5.75~{\rm ms}}\frac{\dot{P}}
{1.99\time 10^{-20}}\right)^{1/20}~{\rm K}~.
\ee
This temperature is consistent with the observed values,
$\sim 1\times 10^6$, only for large multiplication
coefficient, $K\simeq 200$, much greater than the maximum values,
$K_{\rm max}\simeq 1$ and $\simeq 20$ calculated by Bogovalov \& Kotov (1989)
for $\gamma_f=2\times 10^6$ and $10^7$, plausible Lorentz
factors for \psr.
Moreover, 
the same calculations show that $K$ sharply decreases with 
decreasing $B$ from $\simeq
10^{12}$~G and becomes $\ll 1$ at $B\lapr 2\times 10^{11}$~G.
Thus, we have to conclude that either this model strongly overestimates
the PC temperature and luminosity, or the multiplication coefficient
is much greater than calculated.

\subsection{Implications for the magnetic field geometry
and the NS mass and radius}
The fits of the spectra and of the light curves presented in this paper
were obtained for fixed values of the angles $\zeta=40^\circ$ and
$\alpha=35^\circ$ inferred by Manchester \& Johnston (1995)
from the phase dependence of the position angle of the radio
polarization. Since these angles were determined approximately,
and it is not trivial to estimate their uncertainties from
the radio data, we cannot exclude that their true values may be
different. We also adopted canonical values for the NS mass
and radius, $\mns =1.4 M_\odot$ and $\rns =10$~km, whereas the true
mass and radius may differ from these values. We have demonstrated
(see Fig.~7) that the shape of the light curves is very sensitive
to the values of $\zeta$, $\alpha$ and $\mns /\rns$. This opens
a new opportunity to constrain these quantities using them as
free parameters for the light curve fitting. This analysis is
presented elsewhere (Pavlov et al.~1997). Here we only mention
that for {\em any} allowed values of $\alpha$ and $\zeta$, the mass-to-radius
ratio is constrained as $\mns < 1.7 M_\odot (\rns /10~{\rm km})$,
whereas for $\alpha=35^\circ$, $\zeta=40^\circ$ we obtained
$1.4 < (\mns/M_\odot)/(\rns/10~{\rm km}) < 1.6$.
The latter inequality means that if the NS mass equals $1.4 M_\odot$, 
its radius is $8.8 < \rns < 10.0$~km.

\subsection{Implications for the interstellar hydrogen}
We have shown that interpretation of the soft X-ray data in terms
of thermal-like radiation from PCs yields systematically lower
values for the intervening hydrogen column density than the power-law
fits of the spectrum. For instance, BT93 found $n_H=(1.4\pm 0.5)
\times 10^{20}$~cm$^{-2}$ for the one-component power-law fit,
whereas we obtained $n_H\lapr 2\times 10^{19}$~cm$^
{-2}$ and $n_H=(1-5)\times 10^{19}$~cm$^{-2}$ for the single-
and two-temperature PC models, respectively.
It is useful to compare these values with those inferred from
observations of other objects close towards \psr~($l=253^\circ$,
$b=-42^\circ$; $d=180$~pc). The pulsar is in the southern
third galactic quadrant ($180^\circ < l < 270^\circ$, $b<0^\circ$),
apparently within the famous low HI column region that 
extends out to $200-300$~pc at $210^\circ \lapr l \lapr 265^\circ$
(at least for $b\gapr -30^\circ$ --- see, e.~g., Paresce 1984; Welsh 1991).
For instance, two stars of close longitudes and distances, 
$\epsilon$~CMa ($l=240^\circ$, $b=-11^\circ$, $d=187$~pc) and
$\beta$~CMa ($l=226^\circ$, $b=-14^\circ$, $d=220$~pc), show
the neutral hydrogen densities as low as $\log n_{HI}
= 18.0-18.5$ and 18.2--18.4,
respectively (Fruscione et al.~1994; Welsh 1991). The densities in
directions of two other stars within the low density region,
RE 0503--289 ($l=231^\circ$, $b=-35^\circ$, $d=90$~pc)
and RE 0457--281 ($l=229^\circ$, $b=-36^\circ$, $d=90$~pc),
with latitudes closer to that of the pulsar (farther
from the Galactic plane), albeit with twice smaller distances,
are even lower: $\log n_{HI} =17.3-17.8$ and 17.1--17.9.
A white dwarf WD 0320--540 ($l=267^\circ$, $b=-52^\circ$,
$d=103$~pc),
which is likely near the boundary of the low-density
region, shows $\log n_{HI}= 18.9-19.2$. Based on these values
and observations of other (scarce) stars in this region of the
sky (Fruscione et al.~1994, Welsh 1991; and references therein),
we expect that plausible neutral hydrogen column for \psr~ should not exceed
$n_{HI}\simeq (1-3)\times 10^{19}$~cm$^{-2}$.

Interstellar absorption of X-rays enters
the spectral fits through the attenuation function,
$\exp(-n_H\sigma)$, where $\sigma$ is the effective cross section
per hydrogen atom calculated under the assumption 
that the interstellar hydrogen is not ionized (e.~g., Morrison \&
McCammon 1983). The effective hydrogen column density $n_H$
coincides, naturally, with $n_{HI}$ for the nonionized ISM,
but for the partially ionized hydrogen we have
$n_{HI} < n_H < n_H^{\rm tot}$ , where
$n_{H}^{\rm tot}\simeq n_{HI}+n_e$ is the total (neutral plus ionized)
hydrogen column density. With the electron column density
$n_e=0.81\times 10^{19}$~cm$^{-2}$, known from the pulsar dispersion
measure, we can restrict the mean ionization degree
of hydrogen, $\xi=n_e/n_H^{\rm tot}$, as
$n_e/(n_e+n_H) < \xi
< n_e/n_H$. 
If we adopt $n_H=(1-3)\times 10^{19}$~cm$^{-2}$ as a plausible
estimate, the ionization degree 
$0.2<\xi<0.8$
is noticeably greater,
and the (total) mean number density in the direction towards \psr,
$n_H^{\rm tot}/d=
(0.02-0.07) d_{180}^{-1}$~cm$^{-3}$, is lower than the estimates,
$\xi\simeq 0.04-0.1$ and $n_H^{\rm tot}/d\simeq
(0.16-0.34) d_{180}^{-1}$~cm$^{-3}$,
obtained from the power-law fit. The inferred high ionization 
is in line with $\xi\approx 0.9$ found by Gry, York \& Vidal-Madjar (1985)
in the direction to $\beta$ CMa.
It is worth noting that the mean number
density inferred from our PC model fits is fairly close to the ISM density 
of the ambient medium 
around the pulsar estimated by equating the ram pressure
of the interstellar medium to the pulsar wind pressure at the
apex of the pulsar wind's bow shock (see, e.~g., HMM96): $n_H^{\rm tot}=
\dot{E}/(4\pi r_w^2 v_p^2 c m_H) \simeq 0.08$~cm$^{-3}$, where
the distance from the pulsar to the apex of the bow shock,
$r_w=2.4\times 10^{16}$~cm, and the velocity of the pulsar,
$v_p=118$~km s$^{-1}$, are evaluated for $d=180$~pc.

\subsection{X-rays from other millisecond pulsars: thermal or nonthermal?}
The fact that all available X-ray and EUV data are consistent
with our interpretation still does not prove that
this interpretation is unique --- it often happens that the
same data can be equally well fitted with quite different models.
Therefore, it is important to verify that the hypothesis does not
contradict to observations of similar objects. 
According to BT97, soft X-ray radiation has been observed from seven
millisecond pulsars. At least one of them, PSR B1821--24 ($P=3.05$~ms,
$\dot{E}=2.2\times 10^{36}$~erg~s$^{-1}$), 
shows a strong evidence that its radiation is of a nonthermal
(magnetospheric) origin --- its spectrum,
observed with $ASCA$ up to 10~keV, can be fitted with a power law,
and its X-ray pulses are sharp (Saito et al.~1997). Thus, if we
assume that PCs
is a phenomenon common for all millisecond pulsars, we must
explain why this phenomenon is not observed for PSR B1821--24.
This can be done with the use of the aforementioned pulsar models. 
If the correct estimate obtained for the PC luminosity from the slot-gap model
is not a chance coincidence, we may expect that the luminosity
is scaled as $L_{\rm cap} \propto \dot{P}^{1/2} P^{-23/8}$, and
the bolometric luminosity from two PCs of PSR B1821--24 can be
estimated as $L_{\rm bol}\sim (6-9)\times 10^{31}$~erg~s$^{-1}$.
This estimate is much lower than the luminosity
$L_x = 1.7\times 10^{33}$~erg s$^{-1}$ inferred by Saito et al.~(1997).
If we adopt the outer-gap model and assume that the accelerator 
efficiency is reduced by the same factor for all pulsars, then
the luminosity is scaled as $L_{\rm cap} \propto \dot{P}^{1/2}
P^{-11/6}$, which yields $L_{\rm bol}\sim (3-5)\times 10^{31}$~erg~s$^{-1}$
for PSR B1821--24, even less than for the slot-gap model. 
Moreover, even the maximum
possible PC luminosity predicted by the outer-gap model remains lower
than observed. Thus, it is not
surprising that the PC radiation is not seen from this pulsar,
being 
overwhelmed by a nonthermal soft X-ray radiation. Saito
et al.~(1997) speculate that the high luminosity of the nonthermal
radiation may be associated with large values of the magnetic field
at the light cylinder, $B_{\rm lc}\sim 1.5\times 10^6$~G, close to
that of the Crab pulsar and much higher 
than $B_{\rm lc}\sim 1.7\times 10^4$~G for \psr.

The nature of X-ray radiation from 
other five millisecond pulsars 
is much less certain because of small numbers of counts detected.
Their luminosities in the $ROSAT$ range,
inferred under the assumption
of a power-law spectrum with the photon index $\gamma=2$,
obey the same dependence, $L_x\simeq 0.001\dot{E}$,
as ordinary pulsars (BT97), and they are greater than the 
PC luminosities predicted for these objects by the slot-gap model
(scaled to \psr).
However, the ratios of the observed to predicted luminosities,
ranging from $\simeq 1.6$ for PSR J1012+5307 to $\simeq 10$ for
PSR J0751+1907, are 
smaller than $\simeq 25$ for PSR B1821--24.
Since the power-law model always yields a higher luminosity than
the PC model (because of higher $n_H$ needed to suppress
the intrinsically high low-energy flux), 
one may expect that the luminosities
inferred from the PC interpretation would be compatible with those predicted
by equation (3). In addition, as Arons (1981) emphasizes,
equation (3) gives a {\em minimum} expected luminosity ---
although the luminosity of \psr~ coincides with that minimum,
PC luminosities of other pulsars may be greater. Thus, 
the soft X-ray radiation of at least some other millisecond pulsars
can be partly supplied by their PCs, although a nonthermal component
may also be present and may even dominate. 
To resolve the dilemma, and to separate the two
components, further observations are needed. At present, we can
only say that both possibilities should be explored, and 
complementary interpretations are still viable for some pulsars.

\section{Summary and conclusions}

We have shown that both the spectra and the light curves of
the soft X-ray radiation of \psr~
observed with the $ROSAT$ PSPC and HRI and $EUVE$ DSI
can be interpreted as {\em thermal radiation
from two hydrogen-covered
polar caps}. The simplest, single-temperature PC model
allows us to estimate typical PC radius, $\Rpc\sim 1$~km,
and temperature, $\Tpc\sim 1\times 10^6$~K. The
successful fitting of the data
with the two-temperature model indicates that the PC temperature
may be non-uniform, decreasing from $\Tc\sim (1.1-1.8)\times 10^6$~K
at a PC core with a typical size $\Rc\sim 150-400$~m down to
$\Tr\sim (3-5)\times 10^5$~K at a much greater radius
$\Rr\sim 2-6$~km, so that the heated area may comprise a considerable
fraction of the NS surface. The bolometric luminosity of the two PCs is
$L_{\rm bol} = (1.0-1.6)\times 10^{30} d_{180}^2$~erg~s$^{-1}$,
that is $\sim (2-4)\times 10^{-4}$ of the total energy loss of the
pulsar is absorbed and re-emitted
by the PCs.

We emphasize that the proposed interpretation is based essentially
upon the {\em neutron star atmosphere models} --- the data cannot
be interpreted as purely thermal PC radiation with the simplistic
blackbody model. Moreover, with the aid of the NS atmosphere models
we show that the emitting layers should be depleted of heavy
elements, which can be naturally explained by the plausible
assumption that the old NS has experienced accretion of the
hydrogen-rich matter during its long life-time.

The successful fits of the $EUVE$ and $ROSAT$ light curves in different
energy ranges with the PC models are possible only if the  energy-dependent
limb-darkening of the hydrogen atmospheres is taken into account.
The shape of the light curves is also very sensitive to the bending
of photon trajectories in the strong gravitational field of the NS
and to the orientations of the rotational and magnetic axes of the pulsar.
We obtained the satisfactory fits assuming $\mns=1.4 M_\odot$ and $\rns=10$~km,
with the magnetic inclination angle, $\alpha=35^\circ$,
and viewing angle, $\zeta=40^\circ$, inferred from radio observations.
We have demonstrated that the effect
of the gravitational bending on the light curves
can provide important information on the NS mass-to-radius ratio.   

The hydrogen column density
towards the \psr, $n_H\sim (1-3)\times 10^{19}$~cm$^{-2}$,
as well as strong ionization of hydrogen,
$\xi > 20\%$, estimated from the PC model fits are
consistent with the ISM properties obtained  
from observations of other objects in the vicinity of the pulsar,
whereas the power-law fit yields greater $n_H$ and lower $\xi$.
This can be considered as one more argument 
in favor of thermal origin of the X-ray radiation of \psr.

The inferred PC temperature, radius and, especially, luminosity
are in excellent agreement with the predictions of
the slot-gap pulsar model (Arons 1981).
The upper limit on the PC luminosity provided by the
outer-gap pulsar model (Cheng et al.~1986) exceeds the
observed value. This model becomes compatible with our
results if a reduced efficiency of the particle accelerator,
consistent with the upper limit on the gamma-ray flux
from the pulsar, is assumed.

Radio pulsar models predict that heated PCs is  
a common phenomenon inherent to all active pulsars. 
By virtue of this, the dependence $L_x\simeq 0.001 \dot{E}$ found by BT97 for
26 pulsars detected with the $ROSAT$ 
determines a fraction of 
the rotational energy loss re-emitted in X-rays {\em by both nonthermal 
and thermal processes}. Contributions of these two mechanisms to the total 
X-ray flux depend on various factors, such as pulsar period, radius, 
magnetic field, etc.  
In some pulsars (e.~g., PSR B1821--24)
the thermal-like soft X-ray radiation from PCs
may happen to be less intensive than the nonthermal radiation generated
in the pulsar magnetosphere. For instance,
both the slot-gap and outer-gap models predict the luminosity
of PCs of PSR B1821--24 much lower than the luminosity observed with
the $ROSAT$ and $ASCA$. 

Although all the available observations of \psr~ 
are consistent with 
the suggested PC interpretation, and this interpretation is
supported by the theoretical pulsar models and by indirect 
arguments, we still cannot completely exclude that the same data,
collected in the relatively narrow energy range,
could be interpreted with another model ---
for instance, as a combination of the PC radiation and magnetospheric
radiation. 
Hence, the proposed interpretation can 
be considered as complementary to the nonthermal
 interpretation discussed by BT97.
A crucial test 
would be provided by
observations of this object at energies $\gapr 1-2$~keV,
which would enable one to firmly discriminate between the
power-law and thermal-like spectra, or to separate their
contributions. Such observations could be carried out by
$ASCA$ (with a sufficiently long exposure)
and by
the forthcoming $AXAF$, $XMM$ and $ASTRO$-$E$ missions.
Important data could be also obtained from observations
of the pulsar in the UV range ($1200\lapr \lambda\lapr 3000$~\AA)
with the $HST$. In particular, such observations would enable one
to measure the temperature of the entire NS surface 
(cf.~Pavlov, Stringfellow \& C\'ordova 1996a) and to elucidate other possible
heating mechanisms competing with the PC heating. 

For more accurate and reliable interpretation of the future
observations, more theoretical work is highly desirable.
In particular, the temperature distribution over the NS
surface within and around the PCs should be investigated
(and further used to fit the data), and the nonthermal
X-ray spectrum and beam shape should be calculated. 

\acknowledgements
We are grateful to Werner Becker and Jules Halpern for providing us with the 
$ROSAT$ PSPC/HRI and  $EUVE$ DSI
observational data prior to publication. We are greatly indebted to Joachim
Tr\"umper and Werner Becker for their collaboration and
fruitful, stimulating discussions.
V.~E.~Zavlin acknowledges the Max-Planck fellowship.
G.~G.~Pavlov is thankful to the Max-Planck-Institut f\"ur Extraterrestrische
Physik for the warm hospitality.
This work was partially supported through the
INTAS grant 94-3834, DFG-RBRF grant 96-02-00177G and
NASA grant NAG5-2807.


\begin{thebibliography}{}

\bibitem{}
Alcock C., Illarionov A.~F., 1980, ApJ, 235, 534

\bibitem{}
Arons J., 1981, ApJ, 248, 1099

\bibitem{}
Arons J., Scharlemann, E.~T., 1979, ApJ, 231, 854.

\bibitem{}
Bailyn C.~D., 1993, ApJ, 411, L83

\bibitem{}
Becker W., Tr\"umper J. (BT93), 1993, Nat., 365, 528

\bibitem{}
Becker W., Tr\"umper J. (BT97), 1997, A\&A (accepted)

\bibitem{}
Becker W., Tr\"umper J., Brazier K.~T.~S., Belloni T., 1993,
IAU Circular 5701

\bibitem{}
Becker W., et al., 1997 (in preparation)

\bibitem{}
Bell J.~F., Bailes M., Bessell M.~S., 1993, Nature, 364, 603

\bibitem{}
Bell J.~F., Bailes M., Manchester R.~N., Weisberg J.~M.,
Lyne A.~G., 1995, ApJ, 440, L81

\bibitem{}
Beskin V.~S., Gurevich A.~F., Istomin Ya.~N., 1993,
Physics of Pulsar Magnitosphere. Cambridge Univ. Press, Cambridge

\bibitem{}
Bogovalov S.~V., Kotov, Yu.~D. 1989, Sov.~Astron.~Lett., 15, 185
 
\bibitem{}
Bowyer S., et al., 1996, ApJS, 102, 129

\bibitem{}
Cheng A.~F., Ruderman M.~A., 1980, ApJ, 235, 576

\bibitem{}
Cheng K.~S., Ho C., Ruderman M.~A., 1986, ApJ, 300, 500

\bibitem{}
Danziger I.~J., Baade D., Della Valle M., 1993, ApJ, 408, 179

\bibitem{}
Edelstein J., Foster R.~S., Bowyer S., 1995, ApJ, 454, 442

\bibitem{}
Fierro J.~M., et al., 1995, ApJ, 447, 807

\bibitem{}
Fruscione A., Hawkins I., Jelinsky P., Wiercigroch A.,
1994, ApJS, 94, 127

\bibitem{}
Gil J., Krawczyk A., 1997, MNRAS, 285, 561

\bibitem{}
Gry C., York D.~G., Vidal-Madjar A., 1985, ApJ, 296, 593

\bibitem{}
Greiveldinger C., et al., 1996, ApJ, 465, L35

\bibitem{}
Halpern J.~P., Ruderman M., 1993, ApJ, 415, 286

\bibitem{}
Halpern J.~P., Martin C., Marshall H.~L. (HMM96), 1996, ApJ, 462, 908

\bibitem{}
Hernquist L., 1985, MNRAS, 213, 313

\bibitem{}
Johnston S., et al., 1993, Nature, 361, 613

\bibitem{}
Kawai N., Tamura K., Saito Y., 1996, in: Burke W.~R. (ed).
ESA's Report to the 31st COSPAR Meeting 
(ESA-SP 1194). Noordwijk, ESA (in press)

\bibitem{}
Manchester R.~N., Johnston S., 1995, ApJ, 441, L65

\bibitem{}
Manning R.~A., Willmore A.~P., 1994, MNRAS, 266, 635

\bibitem{}
Michel F.~C., 1991, Theory of Neutron Star Magnitospheres.
Univ. of Chicago Press, Chicago

\bibitem{}
Morrison R., McCammon D., 1983, ApJ, 270, 119

\bibitem{}
\"Ogelman H., 1995, in: Alpar M.~A., Kizilo\u{g}lu \"{U}.,  van Paradijs J.
(eds). The Lives of the Neutron Stars. Kluwer, Dordrecht, p.~101

\bibitem{}
Paresce F., 1984, AJ, 89, 1022

\bibitem{}
Pavlov G.~G., Shibanov Yu.~A., 1978, Sov. Astron., 22, 214

\bibitem{}
Pavlov G.~G., Shibanov Yu.~A., Ventura J., Zavlin V.~E., 1994,
A\&A, 289, 837

\bibitem{}
Pavlov G.~G., Shibanov Yu.~A., Zavlin V.~E., Meyer R.~D., 1995,
in: Alpar M.~A., Kizilo\u{g}lu \"{U}., van Paradijs J. (eds).
The Lives of the Neutron Stars. Kluwer, Dordrecht, p.~71

\bibitem{}
Pavlov G.~G., Stringfellow G.~S., C\'ordova F.~A., 1996a,
ApJ, 467, 370

\bibitem{}
Pavlov G.~G., Zavlin V.~E., Becker W., Tr\"umper J., 1996b,
BAAS, 28, 947

\bibitem{}
Pavlov G.~G., Zavlin V.~E., Tr\"umper J., 1997, in:
Olinto A., Frieman J., Schramm D. (eds).
18th Texas Symposium on Relativistic Astrophysics.
World Scientific Press (in press)

\bibitem{}
Radhakrishnan V., Cooke D.~J., 1969, Astrophys. Lett., 3, 225

\bibitem{}
Rajagopal M., Romani R.~W., 1996, ApJ, 461, 327

\bibitem{}
Romani R.~W., 1987, ApJ, 313, 718

\bibitem{}
Romani R.~W., 1996, ApJ, 470, 469

\bibitem{}
Saito Y., et al., 1997, ApJ, 477, L37

\bibitem{}
Sandhu J.~S., et al.,
1997, ApJ, 478, L95 

\bibitem{}
Seward F.~D., Wang Z.-R., 1988, ApJ, 332, 193

\bibitem{}
Shibanov Yu.~A., Zavlin V.~E., Pavlov G.~G., Ventura J., 1995,
in: Alpar M.~A.,  Kizilo\u{g}lu  \"{U}., van Paradijs J. (eds).
The Lives of the Neutron Stars. Kluwer, Dordrecht, p.~91

\bibitem{}
Sturner S.~J., Dermer C.~D., Michel F.~C., 1995, ApJ, 445, 736

\bibitem{}
Wang F.~Y.-H., Halpern J.~P., 1997, ApJ, 482, 159

\bibitem{}
Welsh B.~Y., 1991, ApJ, 373, 556

\bibitem{}
Yancopoulos S., Hamilton T.~T., Helfand D., 1994, ApJ, 429, 832

\bibitem{}
Zavlin V.~E., Shibanov Yu.~A., Pavlov G.~G., 1995,
Astron. Let., 21, 149

\bibitem{}
Zavlin V.~E., Pavlov G.~G., Shibanov Yu.~A. (ZPS96), 1996, A\&A, 315, 141 

\bibitem{}
Zimmermann H.-U., et al., 1994, EXSAS Users's Guide.
MPE Report 257, ROSAT Scientific Data Center, Garching

\end{thebibliography}
\end{document}